\documentclass[prb,preprintnumbers,amsfonts,amssymb,amsmath,floats,twocolumn,aps]{revtex4}

\usepackage[pdftex]{graphicx}
\usepackage{dcolumn}
\usepackage{bm}
\usepackage{color}

\usepackage[pdftex,colorlinks=true,linkcolor=blue,citecolor=blue,filecolor=blue]{hyperref}

\newcommand{\ff}[1]{{\boldsymbol #1}}

\newcommand{\bi}{\begin{itemize}}
\newcommand{\ei}{\end{itemize}}
\newcommand{\be}{\begin{equation}}
\newcommand{\ee}{\end{equation}}
\newcommand{\ba}{\begin{eqnarray}}
\newcommand{\ea}{\end{eqnarray}}

\newcommand{\jd}{J^{\rm dim}_{\rm c}}
\newcommand{\jm}{J^{\rm mag}_{\rm c}}
\newcommand{\jdv}{0.89t}
\newcommand{\jmv}{0.84t}

\begin{document} 
  
\title{Frustrated quantum magnetism in the Kondo lattice on the zigzag ladder}
\author{Matthias Peschke, Roman Rausch and Michael Potthoff}
\affiliation{I. Institut f\"ur Theoretische Physik, Universit\"at Hamburg, Jungiusstra\ss{}e 9, 20355 Hamburg, Germany}

\begin{abstract}
The interplay between Kondo effect, indirect magnetic interaction and geometrical frustration is studied in the Kondo lattice on the one-dimensional zigzag ladder. 
Using the density-matrix renormalization group (DMRG), the ground state and various short- and long-range spin- and density-correlation functions are calculated for the model at half-filling as a function of the antiferromagnetic Kondo interaction down to $J=0.3t$ where $t$ is the nearest-neighbor hopping on the zigzag ladder. 
Geometrical frustration is shown to lead to at least two critical points: 
Starting from the strong-$J$ limit, where almost local Kondo screening dominates and where the system is a nonmagnetic Kondo insulator,  antiferromagnetic correlations between nearest-neighbor and next-nearest-neighbor local spins become stronger and stronger, until at $\jd \approx \jdv$ frustration is alleviated by a spontaneous breaking of translational symmetry and a corresponding transition to a dimerized state. 
This is characterized by antiferromagnetic correlations along the legs and by alternating antiferro- and ferromagnetic correlations on the rungs of the ladder. 
A mechanism of partial Kondo screening that has been suggested for the Kondo lattice on the two-dimensional triangular lattice is not realized in the one-dimensional case.
Furthermore, within the symmetry-broken dimerized state, there is a magnetic transition to a $90^{\circ}$ quantum spin spiral with quasi-long-range order at $\jm \approx \jmv$. 
The quantum-critical point is characterized by a closure of the spin gap (with decreasing $J$) and a divergence of the spin-correlation length and of the spin-structure factor $S(q)$ at wave vector $q=\pi/2$. 
This is opposed to the model on the one-dimensional bipartite chain, which is known to have a finite spin gap for all $J>0$ at half-filling.
\end{abstract} 

\maketitle 

\section{Introduction}
\label{sec:intro}

Kondo lattices \cite{Don77,TSU97b} represent one of the most intensively studied classes of interacting many-body systems in condensed-matter theory and are prototypical systems in the field of quantum magnetism. \cite{Aue94} 
In particular, they are used to discuss the collective magnetic order of heavy-fermion materials \cite{Ste01} with local magnetic moments resulting, e.g., from an incompletely filled $4f$ shell. \cite{DDN98}
The notoriously difficult phase diagrams of the Kondo-lattice model in various dimensions and on lattices with different topologies result from the competition of different fundamental and competing physical mechanisms favoring or suppressing collective magnetic order.
The most prominent ones are the Kondo effect and the RKKY coupling. 

The Kondo effect is already captured by the Kondo impurity model \cite{Kon64} which describes a local magnetic moment antiferromagnetically coupled via a local exchange interaction $J$ to the local spin of a system of itinerant conduction electrons, which hop with amplitude $t$ between the sites of a $D$-dimensional lattice. 
Below a characteristic (Kondo) temperature scale $T_{\rm K} \sim t \, e^{-t/J}$, the local moment is screened by a mesoscopically large cloud of conduction electrons. \cite{Hew93}

While the impurity model is amenable to an exact numerical treatment, \cite{Wil75,BCP08,OKK09a} the Kondo lattice with quantum spins $S=1/2$ coupled via $J$ to the local conduction-electron moments at each lattice site is far from being fully understood.
At weak $J$, not only the emergence of a coherence energy scale, smaller than $T_{\rm K}$, but also indirect magnetic coupling between the spins, mediated via the conduction electrons, such as the RKKY mechanism, \cite{RK54,Kas56,Yos57} complicates the physics.

At half-filling $n=1$ and weak $J$, the emergent RKKY interaction $J_{\text{RKKY}}$ induces nonlocal magnetic correlations which, in case of a bipartite lattice, favors antiferromagnetic long-range magnetic order.
For strong $J \gg t$, on the other hand, local Kondo singlets are formed, and the system becomes a nonmagnetic Kondo insulator with charge and spin gap of the order of $J$ and with residual antiferromagnetic short-range correlations due to an indirect magnetic exchange with a kinetic-energy gain $\sim t^{2}/J$.
As a function of the interaction strength $J$, one thus expects a competition of the nonlocal RKKY coupling and the Kondo effect in the spirit of Doniach. \cite{Don77,ILC97}

While for the one-dimensional lattice ($D=1$) the Mermin-Wagner theorem \cite{MW66} and quantum fluctuations in the ground state exclude a symmetry-broken state, one might still expect a transition to a state with quasi-long-range magnetic correlations when decreasing $J$. 
A zero-temperature quantum-phase transition, however, has been excluded. \cite{TSU97b}
This is opposed to the $D=2$ lattice, where numerically exact quantum Monte-Carlo studies \cite{Ass99,CA01} could demonstrate the existence of a quantum-critical point at $J/t \approx 1.4$ for $n=1$.
Mean-field and cluster mean-field studies \cite{Roz95,ZY00,PP07,MBA10,ABF13,LGM16} and experiments \cite{KS10} support the existence of a phase transition in $D=2$ and higher dimensions.

For a half-filled Kondo lattice on a {\em non-bipartite} lattice, frustration of antiferromagnetic order considerably complicates the physics. 
The magnetically frustrated Kondo lattice is relevant for a couple of materials, e.g., the geometrically frustrated heavy-fermion antiferromagnet  CePdAl on the Kagome-like lattice, \cite{OMN+08} and other frustrated systems. \cite{NIWS93,FHB+04,FTS+06}
From a theoretical perspective, it represents a many-body problem which is highly interesting due to the several competing or cooperating  mechanisms at work which favor or impede magnetic ordering.
Besides $T_{\rm K}$ and $J_{\rm RKKY}$, the energy associated with the release of frustration adds to the problem as a third relevant energy scale.
This may give rise to novel phenomena.

In particular, for the $D=2$ triangular lattice a mechanism of partial Kondo screening (PKS) has been suggested. \cite{MNYU10}
Here, frustration is avoided by a site-selective Kondo-singlet formation such that the remnant moments can order magnetically via the RKKY coupling.
The site-selective cooperation of Kondo screening and RKKY interaction can occur spontaneously or be triggered by the chemical environment.
A possible realization of partial Kondo screening has been studied in the frustrated Kondo lattice and in different model variants, e.g., the Kondo necklace, the Ising-spin Kondo lattice and the Anderson lattice, using different approximations and numerical techniques, including 
Hartree-Fock theory, \cite{HUM11,HUM12a}
Monte-Carlo simulations, \cite{IM12,IM13}
dynamical mean-field theory (DMFT), \cite{PKP11} 
and variational Monte-Carlo. \cite{MNYU10}

Recently, site-selective DMFT has been employed for a systematic study of the competition between indirect exchange and Kondo screening on the triangular lattice. \cite{AAP15} 
The resulting magnetic phase diagram is surprisingly complex and, besides a nonmagnetic Kondo insulator, a heavy-fermion metallic state and antiferro- and ferromagnetic phases, includes an extended parameter region with PKS. 
The PKS phase is in fact exclusively located at the border between the nonmagnetic heavy-fermion and the magnetically ordered phase.
Given the complexity of the problem posed by strong correlated electrons on two-dimensional frustrated lattices, however, the insight gained by site-selective DMFT and also by the previous (approximate) approaches must be questioned seriously.

Quantum Monte-Carlo approaches to the frustrated Kondo lattice in $D=2$ typically suffer from the QMC sign problem. 
For a model variant, however, namely for a half-filled Kondo lattice model on the honeycomb lattice with an additional and geometrically frustrated direct spin-spin interaction, a negative-sign-free auxiliary field QMC algorithm has been developed recently, and a partial-Kondo-screened state is found. \cite{SAG17} 

\begin{figure}[t]
\includegraphics[width=0.75\columnwidth]{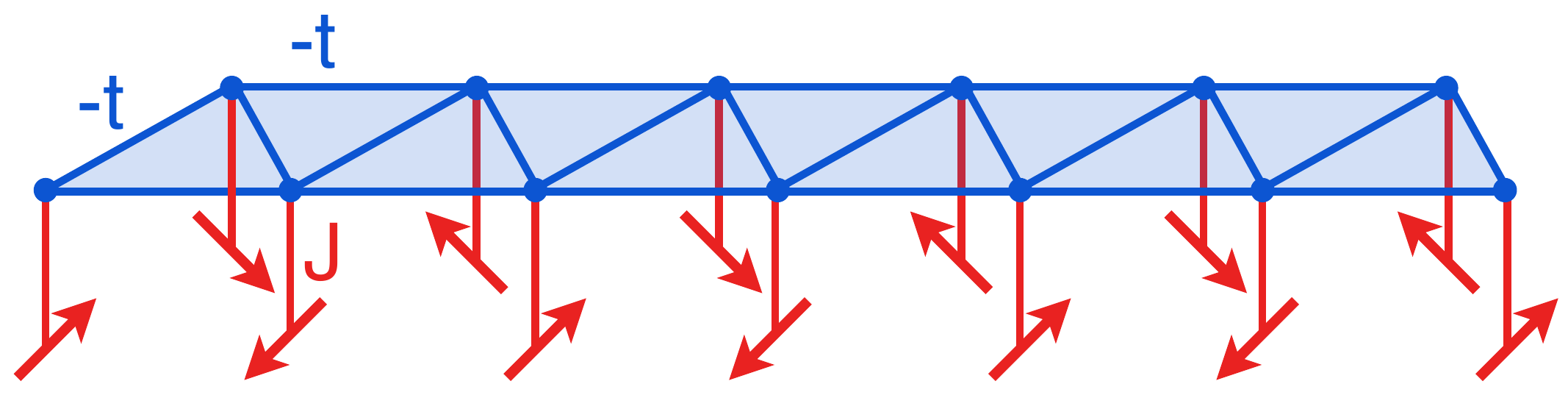}
\caption{
Sketch of the Kondo-lattice model on the one-dimensional zigzag ladder. 
$-t$ denotes the hopping between neighboring sites along the legs and on the rungs.
$J>0$ is the strength of the local antiferromagnetic exchange coupling with the local spin-$1/2$ magnetic moments. 
Energy units are fixed by choosing $t=1$.
}
\label{fig:ladder}
\end{figure}

Here, we tackle the problem along a different route and aim at a numerically exact study of the Kondo lattice on an essentially one-dimensional frustrated lattice, the zigzag ladder (see Fig.\ \ref{fig:ladder}).
Using density-matrix renormalization group (DMRG), \cite{Whi92,Sch11} the phase diagram of the model is studied on lattices with up to $L=60$ sites. 
We employ an implementation of the DMRG which explicitly respects the SU(2) spin-rotation symmetry of the Hamiltonian. \cite{McC07,Wei12} This allows us to reliably study magnetic correlation functions down to interaction strengths $J \approx 0.3t$ at $n=1$. 

Our study focusses on the model at half-filling. 
For a {\em bipartite} lattice (in any dimension), $n=1$ and arbitrary $J$, it is well known \cite{She96,Tsu97a} that the ground state is unique with total spin quantum number $S_{\rm tot} = 0$. 
Spin correlations peak at the antiferromagnetic wave vector $\ff Q = (\pi, \pi, ..)$, reflecting the antiferromagnetic RKKY interaction at weak $J$ as well as the indirect magnetic exchange mechanism at strong $J$.
For a bipartite lattice in $D=1$, in particular, previous exact-diagonalization and DMRG studies \cite{THU92,STU96} in the entire $J$ regime have demonstrated that the ground state is an insulator with gapped spin and charge excitations for all $J>0$. 
Gaps only close at $J_c=0$.
DMRG studies of bipartite two- and $n$-leg Kondo ladders \cite{Xav03,XMD04} support this picture, i.e., a non-magnetic ground state for all $J$.
As mentioned above, this is opposed to the situation in two dimensions where a quantum phase transition is found at a finite coupling strength $J$. \cite{Ass99,CA01} 

For {\em non-bipartite} lattices and for the half-filled Kondo model on the zigzag ladder in particular, the magnetic phase diagram is still unknown.
Only in the strong-coupling limit $J \to \infty$, where the physics is dominated by local Kondo-singlet formation, can one safely expect a nonmagnetic insulator.
This is corroborated by our DMRG calculations. 
For couplings below a critical value $J < \jd \approx \jdv$, however, we find that the local moments on every second rung of the ladder develop strong ferromagnetic nearest-neighbor correlations, thereby alleviating the geometrical frustration and allowing the indirect magnetic coupling to form antiferromagnetic correlations on the remaining rungs and along the legs.
This (electronically and magnetically but not structurally) {\em dimerized} state spontaneously breaks the translation symmetry and must be seen as an alternative to partial Kondo screening. 
In fact, a PKS phase is not observed in the entire $J$ range. 
However, for couplings below a critical interaction $\jm \approx \jmv$, our data demonstrate a transition to a state with quasi-long-range magnetic order, which can be characterized as a 90$^{\circ}$ quantum spin spiral. 
The transition at $\jm$ is marked by a closure of the spin gap for $J \to \jm$ ($J>\jm$) and by a divergence of the spin-structure factor $S(q)$ at wave vector $q=\pi/2$ as well as by a diverging spin-correlation length.
The quantum nature of the magnetic state for $J<\jm$ is evident when comparing with predictions of the classical Heisenberg model on the zigzag ladder (104.5$^{\circ}$ phase). \cite{Dub16} 

A direct experimental test of the predictions is difficult as this would require a realization of the frustrated half-filled Kondo lattice as a quasi-$D=1$ compound (see Refs.\ \onlinecite{TSU97b,DR96}, for example).
Simulations of the model using optical quantum technologies as an ultra cold atomic gas trapped in an optical lattice, however, are well conceivable. \cite{GHG+10,ZBB+14,SHH+14}

The article is organized as follows: 
The next section briefly introduces the model. 
Some details of the method are discussed in Sec.\ \ref{sec:dmrg}. 
Sec.\ \ref{sec:short} and Sec.\ \ref{sec:long} present an extended discussion of the results for the short- and for the long-range spin corrections, respectively. 
A summary of the findings and the conclusions are given in Sec.\ \ref{sec:con}.

\section{Model}
\label{sec:mod}

We study the Kondo-lattice model, i.e., quantum spins with $S=1/2$ coupled to a system of $N$ noninteracting conduction electrons.
The electrons hop with amplitude $-t$ between two-fold spin-degenerate orbitals on neighboring sites of a one-dimensional lattice (zigzag ladder) with $L$ sites, see Fig.\ \ref{fig:ladder}.
$n=N/L$ is the average conduction-electron density which we choose as $n=1$ (half filling). 
Whenever convenient, we set $t \equiv 1$ to fix the energy unit.
The coupling is a local antiferromagnetic exchange of strength $J>0$ between the local spin $\ff S_{i}$ at site $i$ and the local conduction-electron spin $\ff s_{i} = \frac{1}{2} \sum_{\sigma \sigma'} c^{\dagger}_{i\sigma} \ff \tau_{\sigma\sigma'} c_{i\sigma'}$ at the same site. $\ff \tau$ is the vector of Pauli matrices, and $\sigma=\uparrow, \downarrow$.
The Hamiltonian reads
\be
{H} = - t \sum_{\langle i,j \rangle, \sigma} c^{\dagger}_{i\sigma} c_{j\sigma} 
+ 
J \sum_{i} \ff s_{i} \ff S_{i}
\: .
\label{eq:ham}
\ee 
Here, $c_{i\sigma}$ annihilates an electron at site $i=1,...,L$ with spin projection $\sigma$, and $c^{\dagger}_{i\sigma}$ is the corresponding creation operator.
The brackets indicate summation over nearest neighbors in the zigzag geometry (Fig.\ \ref{fig:ladder}).

\begin{figure}
\includegraphics[width=0.75\columnwidth]{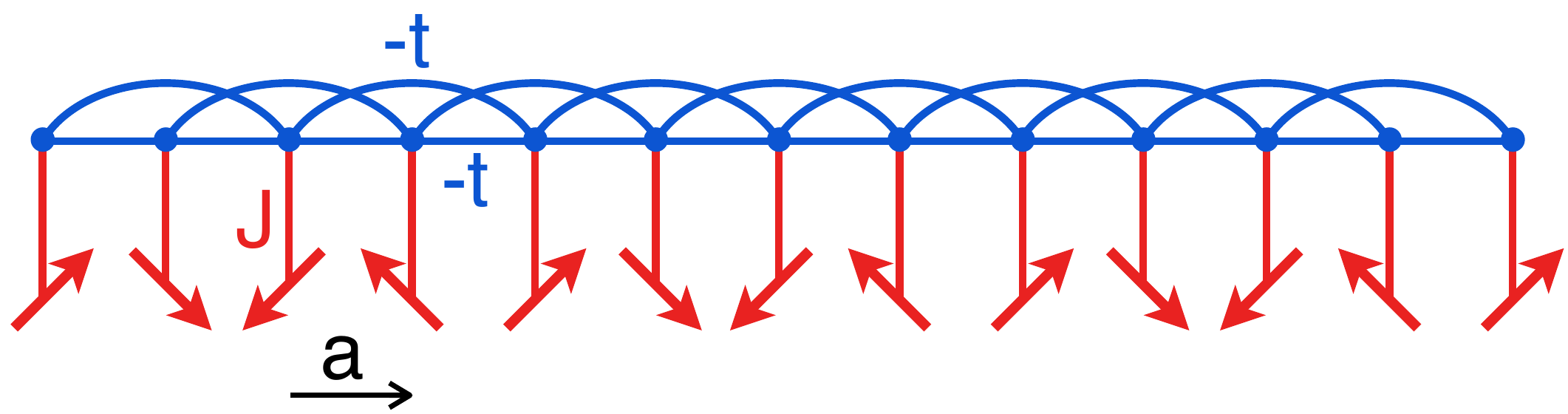}
\caption{
Equivalent sketch of the model, Eq.\ (\ref{eq:ham}), highlighting the translational symmetries. $a$ is the primitive translation.
}
\label{fig:chain}
\end{figure}

The Hamiltonian (\ref{eq:ham}) is invariant under global SU(2) rotations in spin space, generated by the total spin $\ff S_{\rm tot} = \sum_{i} \ff S_{i} + \sum_{i} \ff s_{i}$. 
In the thermodynamical limit, $L\to \infty$, it is furthermore invariant under discrete lattice translations. 
The full translational symmetry group, with the primitive lattice translation $a$, becomes obvious when sketching the model in another way, as done in Fig.\ \ref{fig:chain}. 
Here, the hopping $-t$ connects nearest and next-nearest neighbors of the one-dimensional chain, and the unit cell contains a single site only.
Hopping along the legs is symbolized by curved lines, hopping on the rungs by straight ones.
Note that both representations are fully equivalent.
Due to the finite next-nearest-neighbor hopping the model is not invariant under electron-hole transformations.
Related to this, there is no SU(2) symmetry in the charge sector. \cite{TSU97b}

If periodic boundary conditions were assumed, discrete Fourier transformation of the one-particle basis would diagonalize, for $L<\infty$, the hopping matrix $\ff T$ of the Hamiltonian, i.e., $\ff U^{\dagger} \ff T \ff U = \ff \varepsilon$ with 
\be
  \varepsilon(q) = - 2 t (\cos(qa) + \cos(2qa)) \: ,
\label{eq:dispersion}
\ee  
resulting in a noninteracting band width of $W=6.25t$, and where
\be
  U_{R_{j},q}
  =
  \frac{1}{\sqrt{L}}
  e^{i q R_{j}} 
\label{eq:trafo}  
\ee
with $R_{j} = j a = 0, \pm a, \pm 2a, ...$ is the unitary transformation matrix. 
This transformation also defines correlation functions in reciprocal space, e.g., the spin-structure factor
\be
  S(q) = \frac{1}{L} \sum_{ij}^L e^{i q (R_{i}-R_{j})} 
  \langle \ff S_{i} \ff S_{j} \rangle \: 
\label{eq:scorr}  
\ee
in the ground state $| 0 \rangle$.
For technical reasons, most DMRG computations are performed for the system assuming {\em open} boundaries. 
Spin-structure factors can still be calculated via Eq.\ (\ref{eq:scorr}) for large $L$ when carefully controlling finite-size effects.

\section{Density-matrix renormalization group}
\label{sec:dmrg}

State-of-the-art density-matrix renormalization group (DMRG) techniques in the language of matrix-product states and operators \cite{Whi92,Sch11} are used to compute the ground state, ground-state expectation values and correlation functions.
Both variants of representing the model, the zigzag geometry with $L/2$ unit cells and 2 sites per cell (Fig.\ \ref{fig:ladder}) and the chain geometry with $L$ unit cells containing a single site only (Fig.\ \ref{fig:chain}), have been implemented and tested against each other. 
Due to the smaller local Hilbert space and despite the longer-ranged hopping, the chain geometry is clearly favorable for practical DMRG computations.

Studies of the weak-$J$ regime are computationally demanding due to a comparatively high entanglement entropy which we attribute to the emergence of longer-ranged RKKY effective interactions.
Furthermore, due to the competition between Kondo screening, RKKY coupling and frustration, a complex ground-state ``landscape'' can be expected on relevant energy scales decreasing with decreasing $J$. 
Still we have been able to achieve well-converged results for couplings larger than $J = 0.3t$ at $L=40$ electronic sites and the same number of local spins $\ff S_{i}$. 
For $L=60$ reliable calculations can be performed down to $J=0.7t$.
The overlap between matrix-product states after a half sweep through the lattice differs from unity by less than $1 \times 10^{-7}$.

\begin{figure*}
\includegraphics[width=1.7\columnwidth]{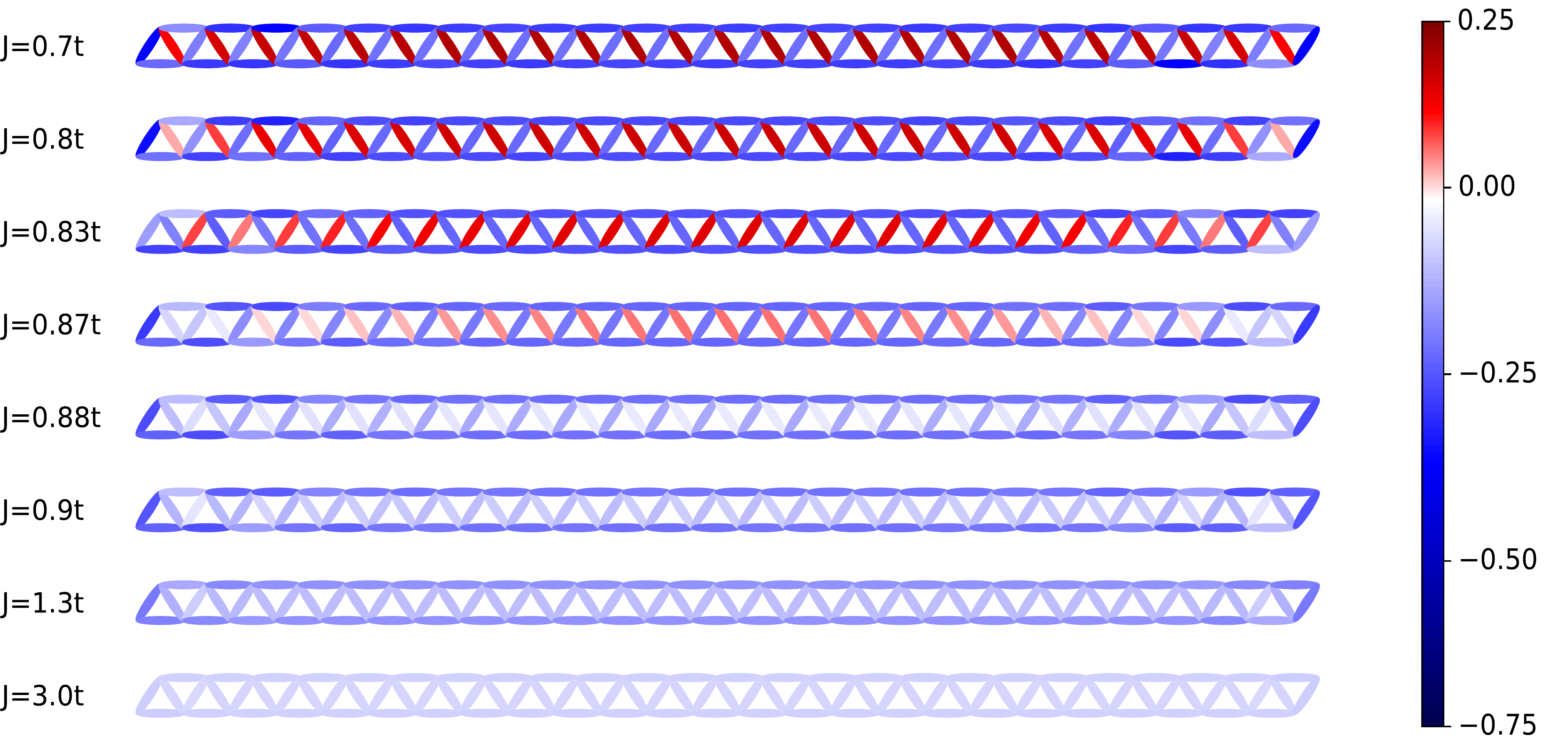}
\caption{
Spin-correlation function $\langle \ff S_{i} \ff S_{j} \rangle$ (see color code) for nearest neighbors $i, j$ on the zigzag ladder as obtained for the ground state of the Kondo lattice with $L=52$ sites at half-filling and for various coupling strengths $J/t$.
}
\label{fig:shortj}
\end{figure*}

Exploiting the symmetries of the Hamiltonian is crucial for an efficient computation, in particular close to the critical interaction $\jm \approx \jmv$ (see below).
Besides the standard U(1) gauge symmetry in the charge sector, our implementation explicitly respects the non-Abelian SU(2) spin-rotation symmetry as well.
Here, we follow previous work described in Refs.\ \onlinecite{McC07,Wei12}.

For the present study, we keep up to $m = \sum_{\alpha} n_{\alpha} \approx 8,000$ density-matrix eigenstates.
These are grouped into blocks labeled by $\alpha = ( N , S )$, i.e., by the irreducible representations of U(1) and SU(2), respectively. 
Each block consists of $d_{\alpha} = 2S+1$ identical $n_{\alpha} \times n_{\alpha}$ subblocks where typically $n_{\alpha} \lesssim 200$.
If only abelian symmetries were exploited, the according number of states kept would be given by 
$m = \sum_{\alpha} d_{\alpha} n_{\alpha} \approx 40,000$ states.

In the strong-$J$ limit, we also checked against results obtained by the recently suggested variationally uniform matrix-product-state approach (VUMPS). \cite{ZSVF+17} 
Here, the main idea is to eliminate the unwanted effects of the open boundary conditions, which are typically assumed in conventional DMRG calculations. 
The VUMPS approach respects the translational symmetries of the infinite system.

\section{Short-range spin correlations}
\label{sec:short}

Contrary to the one-dimensional half-filled bipartite Kondo lattice, \cite{TSU97b} the phase diagram for the zigzag ladder is characterized by at least two critical points at finite interaction strengths. 
Starting from the strong-$J$ limit, the first one marks the transition to a dimerized phase and is located at $J = \jd \approx \jdv$.
This can be determined rather precisely by analyzing the short-range correlations between the local spins, i.e., $\langle \ff S_{i} \ff S_{j} \rangle$ for neighboring sites $i, j$ on the zigzag ladder (corresponding to nearest- and next-nearest neighbors in the chain geometry).
Fig.\ \ref{fig:shortj} displays the short-range spin correlations for a system with $L=52$ sites and different $J$.

\subsection{Strong Kondo coupling}

In the limit $J \to \infty$ the system is a nonmagnetic Kondo insulator with antiferromagnetic Kondo correlations $\langle \ff s_{i} \ff S_{i} \rangle \to -3/4$ (not shown) and very weak nearest-neighbor correlations between the local spins $\langle \ff S_{i} \ff S_{j} \rangle \to 0$.
At finite but strong $J$ (see, e.g., $J=3.0t$ in Fig.\ \ref{fig:shortj}), the local spins start to develop weak antiferromagnetic correlations, 
$\langle \ff S_{i} \ff S_{j} \rangle \approx -0.05$.
Here, the (almost local) Kondo screening dominates over the nonlocal indirect magnetic exchange.
The predominantly local character of the ground state reflects itself in the almost complete absence of boundary effects, i.e., the spin correlation are nearly homogeneous, despite the fact that the lattice is cut at $i=1$ and $i=L$.

That the local spins on each bond (weakly) correlate antiferromagnetically, is easily traced back to an indirect magnetic exchange mechanism which is reminiscent of Anderson's superexchange but with the important difference that the unperturbed ground state is nondegenerate:\cite{And79,TSU97b}
Only if the local spins are aligned antiferromagnetically can an electron virtually hop to the neighboring site and back and thereby gain kinetic energy and thus lower the system's total energy.
Otherwise, for ferromagnetic alignment of the local spins, the conduction-electron spins on neighboring sites are ferromagnetically aligned as well (since $J \gg t$), and the hopping process is blocked by the Pauli principle. 
This well-known argument applies to neighbors along the legs and along the rungs of the ladder; both are linked by the hopping term.
Comparing with a Kondo lattice on the standard bipartite chain (or, referring to Fig.\ \ref{fig:chain}, with vanishing next-nearest-neighbor hopping), the antiferromagnetic correlation is weaker in absolute magnitude. 
This is an expected effect of the geometrical frustration. 

\subsection{Inhomogeneities}

For weaker $J$ (see, e.g., $J=1.3t$ and $J=0.9t$ in Fig.\ \ref{fig:shortj}) inhomogeneities in the correlation function are apparent.
One must distinguish between two effects: 
First, there is a clearly visible boundary effect which originates from the different coordination numbers $z_{i}$ of sites at and close to the edges, i.e., $z_{1}=z_{L}=2$, $z_{2}=z_{L-1}=3$, as compared to the rest of the sites with coordination $z_{i} = 4$.  
Roughly, a lowered coordination implies less frustration and therewith explains that the spin correlation is significantly stronger antiferromagnetic on the first and on the last bond [$(1,2)$ and $(L-1,L)$].
For $J=1.3t$, this unwanted boundary effect only slightly extends into the bulk of the system, while for $J=0.9t$, which is already close to $\jd$, a nearly homogeneous spin correlation is obtained not earlier than at a distance $i \gtrsim 15$ ($i \lesssim L-15$) from the edges.

We have checked the DMRG results displayed in Fig.\ \ref{fig:shortj} against the variationally uniform matrix-product-state approach (VUMPS), \cite{ZSVF+17} which enforces homogeneous correlation functions. 
Perfect agreement with the bulk nearest-neighbor correlations, as obtained with the conventional DMRG, is found for interactions of $J \gtrsim 1.5t$.
For weaker $J$ and thus for smaller charge gaps, however, the VUMPS approach, which does not fix the total particle number, becomes too expensive computationally due to the necessity to adjust the chemical potential.

The second effect is the apparently stronger antiferromagnetic correlation on the legs of the ladder as compared to the rungs, see (see $J=1.3t$ and $J=0.9t$ in Fig.\ \ref{fig:shortj}).
This is a bulk effect and specific to the ladder geometry. 
Homogeneous short-range correlations must be expected for a two-dimensional triangular lattice as all nearest-neighbor bonds are equivalent. 
Here, for the zigzag ladder, 
frustration of the antiferromagnetic interaction between two neighboring sites $i$ and $j$ on a leg comes into play by a {\em single} path involving two nearest-neighbor hops, $i \to k$ and $k \to j$. 
For two neighboring sites $i$ and $k$ on a rung, however, there are {\em two} such paths, each involving two nearest-neighbor hops, $i \to j$ and $j \to k$ and $i \to j'$ and $j' \to k$. 
Hence, antiferromagnetic alignment of neighboring spins on a leg is less frustrated and the absolute magnitude of the correlation function is larger, as can be seen in the figure. 
Comparing the results for $J=1.3t$ with those of $J=0.9t$, the frustration effect and therewith the inhomogeneity grows.
The reason is that, quite generally, the nonlocal indirect magnetic coupling becomes more important with decreasing $J$ at the expense of a decreasing local Kondo correlation $\langle \ff s_{i} \ff S_{i} \rangle$.
 
\subsection{Dimerization}

\begin{figure}[t]
\includegraphics[width=0.95\columnwidth]{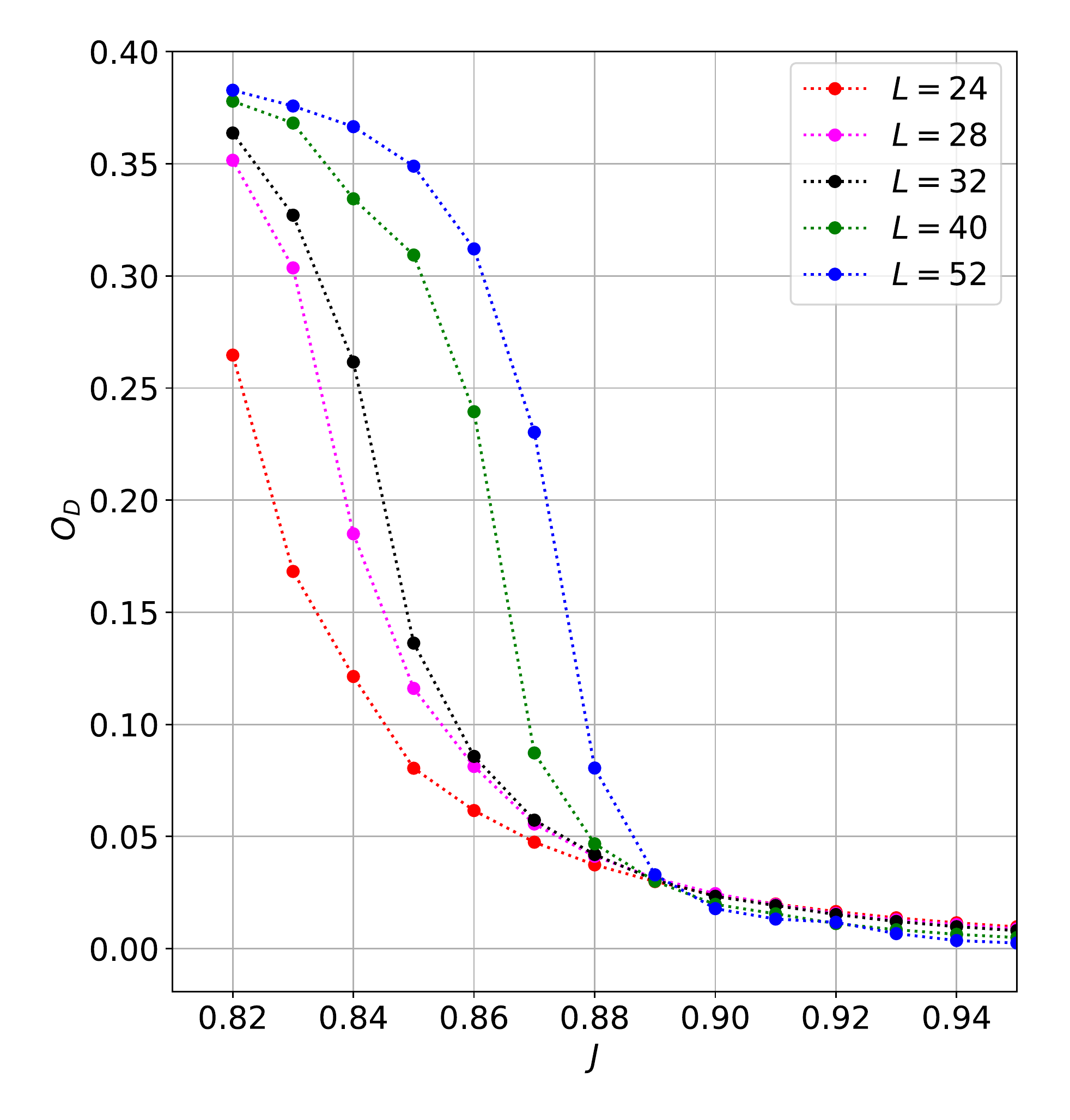}
\caption{
$J$-dependence of the order parameter for dimerization $O_{\rm D}$, Eq.\ (\ref{eq:od}), in the critical $J$-range for different $L$.
}
\label{fig:jcd}
\end{figure}

\begin{figure*}
\includegraphics[width=1.95\columnwidth]{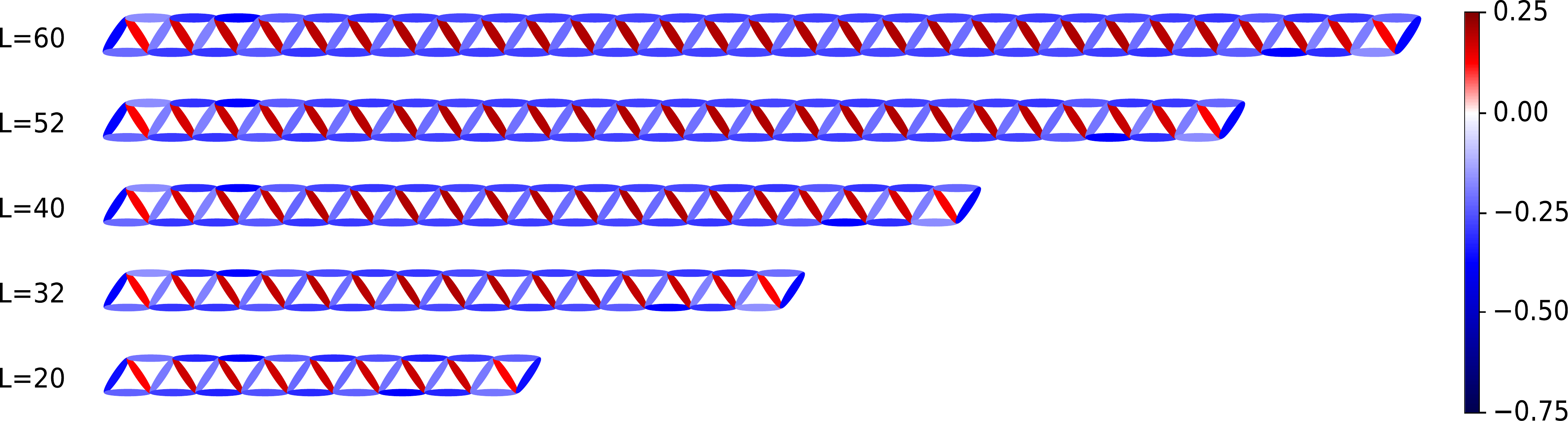}
\caption{
Nearest-neighbor spin correlations as in Fig.\ \ref{fig:shortj} but for different $L$ at fixed $J=0.7t$.
}
\label{fig:shortl}
\end{figure*}

At the critical interaction $\jd \approx \jdv$ we observe another kind of inhomogeneity (in the bulk of the system), namely correlations start to differ on different rungs. 
Fig.\ \ref{fig:shortj} shows that for $J\le \jd$ the spin correlation $\langle \ff S_{i} \ff S_{j} \rangle$ is weaker for $\diagdown$-rungs as compared to $\diagup$-rungs.
With decreasing $J$, the $\diagdown$-correlations decrease in absolute magnitude, vanish for $J \approx 0.875t$ and get positive, i.e., ferromagnetic for $J = 0.87t$ and become even stronger ferromagnetic for $J=0.83t$. 
Let us add that one finds the same ``dimerization pattern'' also in the conduction-electron system namely, e.g., in the spin-spin correlation function $\langle \ff s_{i} \ff s_{j} \rangle$ for $J<\jd$.
 
For $L\to \infty$, $\diagdown$-rungs and $\diagup$-rungs are geometrically equivalent. 
Therefore, the unequal spin (and density) correlations on the two types of rungs must be seen as a spontaneous breaking of the translation symmetry of the ground state.
The dimerized state is invariant only under translations of the reduced translation-symmetry group $\langle 2a \rangle = \{ {\rm id}, \pm 2a, \pm 4a, ... \}$, which is generated by $2a$, rather than by the primitive lattice translation $a$ (here we refer to the chain geometry, Fig.\ \ref{fig:chain}).

Due to the presence of the boundaries in the finite system, however, the rungs are in fact not completely equivalent, and ferromagnetic $\diagdown$-bonds are preferred over ferromagnetic $\diagup$-bonds. 
Namely, for the specific system chosen here, starting and terminating with a $\diagup$-bond, the total number of $\diagup$-bonds is greater by one bond, and thus the total number of ferromagnetic bonds is minimized if the $\diagdown$-bonds are ferromagnetic.
We conclude that the dimerization is triggered by the system's boundaries. 
For large systems, however, the relative energy difference $\Delta E_{\rm tot} / E_{\rm tot}$ of the two corresponding nearly degenerate ground states becomes so small that the DMRG algorithm can be trapped in a metastable ``wrong ground state'', depending on the initial state chosen at the start of the algorithm and on the effect of the fluctuations implemented for an improved convergence.
An example can be seen for $J=0.83t$ in Fig.\ \ref{fig:shortj}.

A rather precise value for the critical interaction for the dimerization can be obtained by analyzing the dimerization order parameter
\be
  O_{\rm D} 
  = 
  \langle \ff S_{i} \ff S_{j} \rangle_{(i,j)=\diagdown}
  - 
  \langle \ff S_{i} \ff S_{j} \rangle_{(i,j)=\diagup} 
  \: ,
\label{eq:od}  
\ee
which is nonzero in the dimerized state. 
Fig.\ \ref{fig:jcd} displays the $J$-dependence of $O_{\rm D}$ close to $\jd \approx \jdv$.
Data for $O_{\rm D}$ have been taken in the vicinity of the center of the system and averaged for equivalent bonds within the bulk region where boundary effects can be neglected.
(Note that $O_{\rm D} = O_{\rm D, i}$ depends on $i$ and shows an alternating sign $(-1)^{i}$).
For $J=0.90t>\jd$, the order parameter monotonically decreases with increasing system size $L$ while for $J=0.88t<\jd$, it monotonically increases.
While extrapolation to $L=\infty$ is not yet reasonable in this small parameter range and would require results for still larger $L$, one can safely conclude that $0.885t < \jd < 0.895t$.
Our data are consistent with either a continuous behavior of $O_{\rm D}(J)$ at $\jd$ or with a transition that is weakly first order.

\subsection{Boundary effects}

Boundary effects are analyzed in the symmetry-broken state at $J=0.7t$ by comparing calculations for different chain lengths $L$, ranging from $L=20$ to $L=60$, see Fig.\ \ref{fig:shortl}. 
Even for the shortest chain, $L=20$, the nearest-neighbor correlations of the local spins appear perfectly homogeneous in the chain center and are almost the same as for the larger systems. 
Independent of $L$, boundary effects are seen up to distances $i \lesssim 5$ ($i\gtrsim L - 5$) from the edges -- as can judged from the color-code presentation.
One should note, however, that the strength of boundary effects depend on $J$. 
They are most pronounced close to the critical interaction $\jd$, as can be seen in Fig.\ \ref{fig:shortj}.
Furthermore, the data do exhibit some Friedel-like oscillations with small and decaying amplitudes which are present up to longer distances (not visible in Fig.\ \ref{fig:shortl}), but this does not change the interpretation of the results.
The wavelength of these oscillations is given by $\lambda_{\rm Friedel} = 2\pi / Q$, where $Q$ is the nesting vector connecting the closest Fermi points in the noninteracting band, see Eq.\ (\ref{eq:dispersion}). 
At half-filling, there are four (spin-degenerate) Fermi points, and the nesting vector is $Q=\pm \pi / 2$, which is consistent with our data.

\subsection{Weak Kondo coupling}

At $J=0.7t$ (see Fig.\ \ref{fig:shortj}) the ferromagnetic correlation on the $\diagdown$-bond is already large, $\langle \ff S_{i} \ff S_{j} \rangle \approx 0.201$, but it further increases with decreasing $J$.
For $J=0.3$ (not shown) we find $\langle \ff S_{i} \ff S_{j} \rangle \approx 0.245$, and thus the correlation is close to its maximum.
Calculations in this parameter regime become excessively time consuming, and converged results have been obtained down to $J=0.3t$ only with the help of the SU(2)-symmetric DMRG code. 

For even weaker $J$, we have not been able to reach full convergence as a function of the DMRG bond dimension.
Still, there is some evidence that around $J=0.2t$ the ground state is, apart from boundary effects, at least close to a valence-bond solid with perfect {\em antiferromagnetic} singlets, $\langle \ff S_{i} \ff S_{j} \rangle = -0.75$, on the $\diagdown$-bonds and vanishing spin correlations else.
For $L\to \infty$, one would expect degenerate $\diagdown$ and $\diagup$ valence-bond states. 
Of course, a valence-bond solid represents another conceivable and interesting trade-off to deal with the geometrical frustration.
In fact, we expect different magnetic structures emerging in the very weak coupling regime $J \ll 0.3t$.
To make firm statements, however, requires more accurate computations.

Opposed to the correlations between the local spins, the $J$-dependence of the local Kondo correlations $\langle \ff s_{i} \ff S_{i} \rangle$ is rather featureless.
Starting off with the minimal value $\langle \ff s_{i} \ff S_{i} \rangle \to -3/4$ for $J\to \infty$, the bulk Kondo correlation is still strong at $J=3.0t$, for instance, where $\langle \ff s_{i} \ff S_{i} \rangle \approx - 0.62$. 
Its modulus monotonically decreases with decreasing $J$, and for $J=0.9t$ we find $\langle \ff s_{i} \ff S_{i} \rangle \approx - 0.23$. 
The Kondo correlation is basically unaffected by the transition at $\jd$.
Interestingly, it also stays perfectly homogeneous, even for $J<\jd$ in the state with reduced translational symmetry, except for deviations from the bulk value for sites $i$ close to one of the edges.
For example, at $i=1$, $\langle \ff s_{1} \ff S_{1} \rangle \approx - 0.67$ at $J=3.0t$, and $\langle \ff s_{1} \ff S_{1} \rangle \approx - 0.34$ at $J=0.9t$, i.e., particularly for weaker $J$ the system boundaries can manifest themselves quite strongly in the Kondo correlations.

Homogeneity of the Kondo correlations in the bulk of the system, however, excludes partial Kondo screening (PKS) as a mechanism to alleviate frustration. 
Instead, this is achieved with the dimerization mechanism as described above.
An extension of the dimerization mechanism to the two-dimensional triangular lattice is conceivable, e.g., one-dimensional chains with nearest-neighbor antiferromagnetic correlations but with ferromagnetic correlations between adjacent chains.
Checking this or other variants is, however, beyond our present-day capabilities, see the discussion in Ref.\ \onlinecite{AAP15}, for example. 

In the limit $J \to 0$, perturbative RKKY theory applies. \cite{RK54,Kas56,Yos57} 
We can compute the static, nonlocal magnetic susceptibility $\chi_{ij}(\omega=0)$ of the conduction-electron system for $J=0$ as the linear response at site $i$ to a local magnetic field applied to the system at site $j$ ($H \mapsto H - \sum_{i} \ff B_{i} \ff s_{i}$): 
\be 
  \chi_{ij}(\omega=0)
  = 
  \frac
  {\partial \langle s_{iz} \rangle}
  {\partial B_{jz}} 
  \Bigg|_{B_{jz}=0}
  =
  \lim_{\beta \to \infty} \int_{0}^{\beta} d\tau \, \langle s_{iz}(\tau) s_{jz}(0) \rangle
  \: .
\ee
Here, $B_{jz}$ is the $z$-component of the field at $j$, and $s_{iz}(\tau) = e^{H_{\rm e} \tau} s_{iz}e^{-H_{\rm e} \tau}$, where $H_{\rm e}$ is the hopping part of the Hamiltonian (\ref{eq:ham}).
With this at hand, the effective RKKY coupling is given by $J^{\rm RKKY}_{ij} = - J^{2} \chi_{ij}(\omega=0)$ and $H_{\rm RKKY} = \sum_{ij} J^{\rm RKKY}_{ij} \ff S_{i} \ff S_{j}$.
Numerical computations are easily done for a finite system, see Ref.\ \onlinecite{SGP12}.
For $L=52$ we find $J^{\rm RKKY}_1/t = + 0.003 \pm 0.002$ (antiferromagnetic) and $J^{\rm RKKY}_{2}/t = - 0.010 \pm 0.004$ (ferromagnetic) for the nearest-neighbor (rungs) and the next-nearest-neighbor (legs) interaction, respectively.
The ``errors'' indicate the standard deviation of the couplings across the chain. 
One may also compare with the values $J^{\rm RKKY}_{1}/t \approx 0.004$ and $J^{\rm RKKY}_{2}/t \approx - 0.009$ obtained for the same system but with periodic boundary conditions. 
In any case, the RKKY couplings are clearly at variance with the DMRG results for the respective short-range spin correlations at the lowest accessible coupling $J=0.3t$. 
In particular, $J^{\rm RKKY}_{1,2}$ are not frustrated within RKKY theory at all. 

We conclude that there are effectively antiferromagnetic interactions in the $J$ range that spans from the superexchange regime (strong $J$) at least down to $J \approx 0.3t$, where a nonperturbative indirect magnetic exchange is at work, different from RKKY.
Contrary, non-frustrated ferromagnetic legs and antiferromagnetic rungs are obtained within the RKKY approach, which we expect to be applicable for $J \ll 0.3t$.
One should also note that, in one spatial dimension, the RKKY interactions $J^{\rm RKKY}_{ij}$ decay very slowly with increasing $|i-j|$ (see Ref.\ \onlinecite{TSU97b}).
This also implies that our DMRG data for the Kondo lattice on the zigzag ladder cannot be interpreted within the framework of a much simpler $J_{1}$-$J_{2}$ Heisenberg ladder, which is explored very well, see Ref.\ \onlinecite{KSSR10} and references therein.

\section{Quasi-long-range order}
\label{sec:long}

The reduced translational symmetry for $J<\jd \approx \jdv$ that has been seen in the short-range spin correlations actually accompanies a magnetic phase transition at $\jm \approx \jmv$. 
For $J<\jm$ the system develops a $90^{\circ}$ quantum spin-spiral.
Noncollinear quasi-long-range magnetic order is indicated by the divergence of the spin-structure factor with increasing system size $L$ at the wave vectors $Q = \pm \pi/2$ in reciprocal space.
For $J>\jm$ the system is a correlated insulator with frustrated short-range antiferromagnetic effective couplings and short-range spin correlations, which are reminiscent of a {\em classical} frustrated spin system.

\subsection{Spin-correlations in real space}

\begin{figure}
\includegraphics[width=0.98\columnwidth]{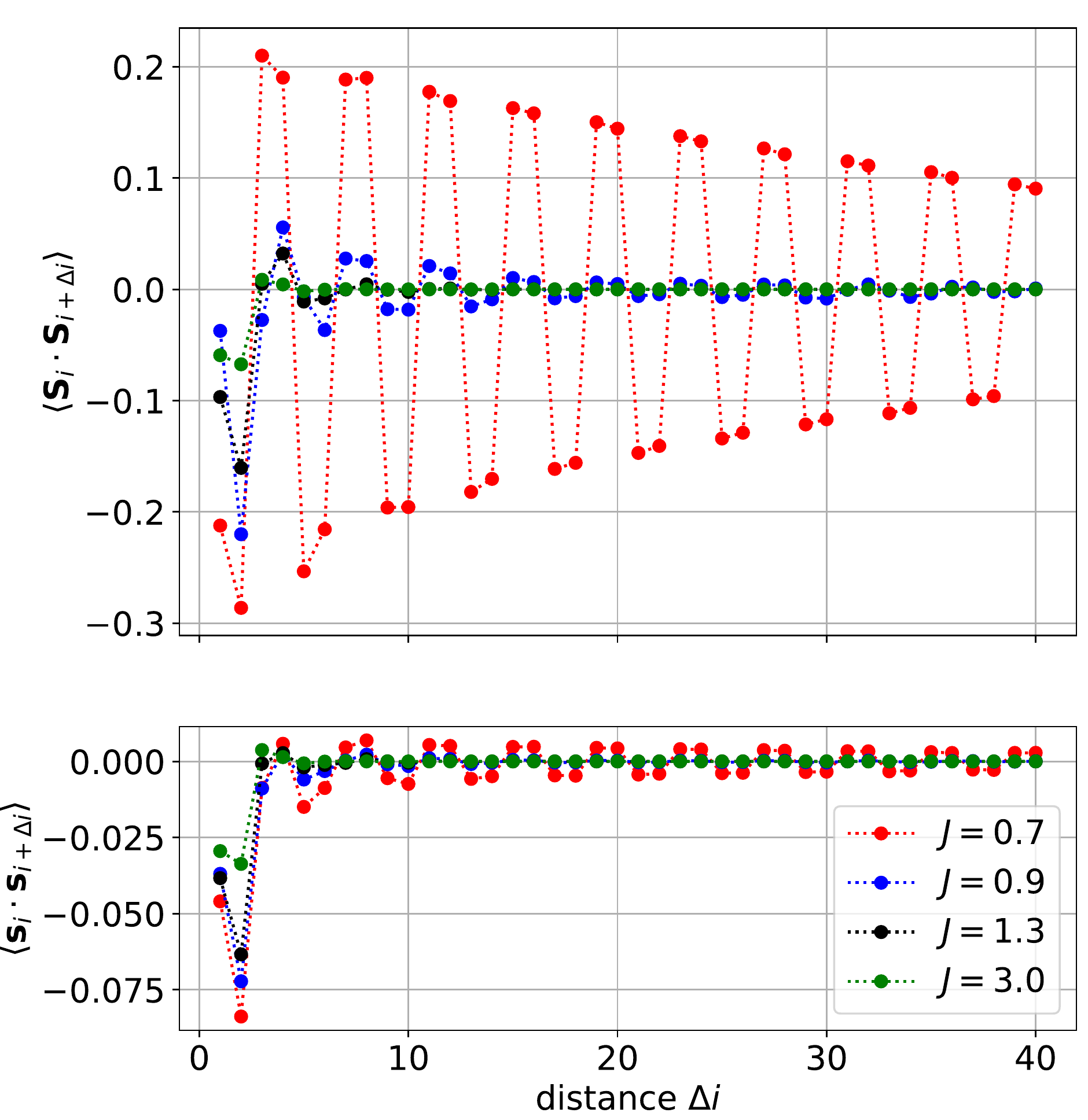}
\caption{
Top: Spin correlation function $\langle \ff S_{i} \ff S_{j} \rangle$ for $i=10$ as a function of the distance $\Delta i \equiv j-i$, obtained for the ground state of the Kondo lattice on the zigzag ladder with $L=60$ sites at half-filling and for various coupling strengths $J$ (with $t=1$).
Bottom: the same but for the conduction-electron spin correlations $\langle \ff s_{i} \ff s_{i+\Delta i} \rangle$.
}
\label{fig:distance}
\end{figure}

To corroborate these propositions, consider first Fig.\ \ref{fig:distance} (top panel) which displays the correlation of the local spins $\langle \ff S_{i} \ff S_{i+\Delta i} \rangle$ in the bulk of the system with $L=60$ as a function of the distance $\Delta i$ between the spins.
Choosing $i=10$ as a test point ensures that boundary effects are invisible on the scale of the figure.
We note that the nearest- and the next-nearest-neighbor spin correlation ($\Delta i=1$ and $\Delta i =2$) are antiferromagnetic for all $J$, above and below $\jm \approx \jmv$.
For $J>\jm$ and with increasing distance $\Delta i$, the spin correlations quickly decay, while for $J = 0.7 t < \jm$ we find the spin correlations to basically extend over the entire system and to decay very slowly.
This behavior is consistent with our expectations for a magnetic phase transition in a one-dimensional system and the onset of quasi-long-range order characterized by a change of the decay of spin correlations from fast exponential to slow algebraic decay.
We also note that with decreasing $J$ (eventually approaching the RKKY regime), the decrease of the local Kondo correlations (discussed in Sec.\ \ref{sec:short}) is consistent with the fact that the localized-spin system becomes more and more decoupled on the low-energy scale and thus develops stronger and more nonlocal correlations. 

The bottom panel of the figure displays the spin correlations of the conduction-electron system $\langle \ff s_{i} \ff s_{i+\Delta i} \rangle$.
Their behavior is found to be qualitatively the same as for the localized spins, in particular the nearest- and the next-nearest-neighbor correlations are negative for all $J$ again, and for large distances correlations decay very slowly when $J< \jm$. 
Also the oscillations in the distance-dependence of the correlations are very similar. 
This is reasonable since the correlations of the local spins are actually mediated by the conduction-electron system.
The absolute magnitude of the conduction-electron spin correlations, however, is considerably smaller than the local-spin correlations.
This can be attributed to the strong delocalization of the conduction electrons in the considered $J$-range. 
A measure for this is the local spin moment which is found to be almost site-independent.
At $J=0.7t$, for instance, it amounts to $\langle \ff s_i^{2} \rangle \approx 0.4$, i.e., the local moment is not well developed and is only slightly larger than the noninteracting Fermi-gas value $\langle \ff s_i^{2} \rangle_{0} = 3/8$.

\subsection{Spin-structure factor}

\begin{figure}[t]
\includegraphics[width=0.8\columnwidth]{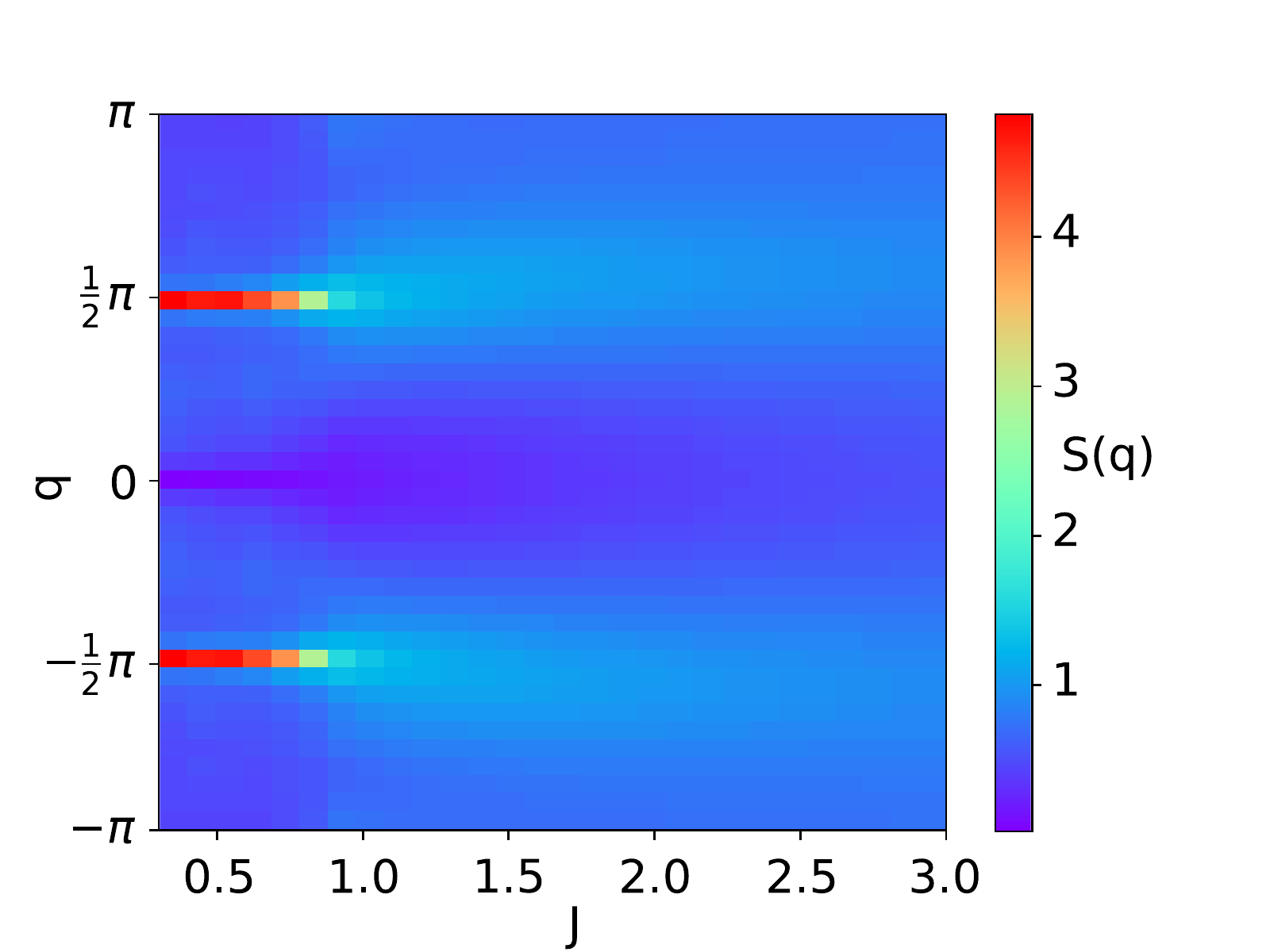}
\caption{
$J$-dependence of the spin-structure factor $S(q) = L^{-1} \sum_{i,j}^L e^{i q (R_{i}-R_{j})} \langle \ff S_{i} \ff S_{j} \rangle$
(color code on the right) for $L=40$ sites.
}
\label{fig:overview}
\end{figure}

The wavelength of the oscillations in the $\Delta i$-dependence of the spin correlations for $J = 0.7 t < \jm$ is given by $\Delta i = 4$. 
This implies that the system, due to geometrical frustration, develops a quantum spin-spiral state with wave vectors $Q = \pm \pi /2$. 
We have performed calculations for the system with $L=40$ and various $J$. 
Fourier transformation of the data, see Eq.\ (\ref{eq:scorr}), yields the spin-structure factor, i.e., the spin-spin correlation function in reciprocal space $S(q)$. 
Fig.\ \ref{fig:overview} provides an overview over $S(q)$ in a wide $J$-range. 
For $J\gg \jm \approx \jmv$ we find a nearly featureless behavior, i.e., long-range spin correlations are not well developed. 
Contrary, for $J < \jm$ one observes a strong enhancement of $S(q)$ at $Q=\pm \pi/2$, which we expect to develop into a divergence when approaching the thermodynamic limit $L\to \infty$.

\begin{figure}[b]
\includegraphics[width=0.8\columnwidth]{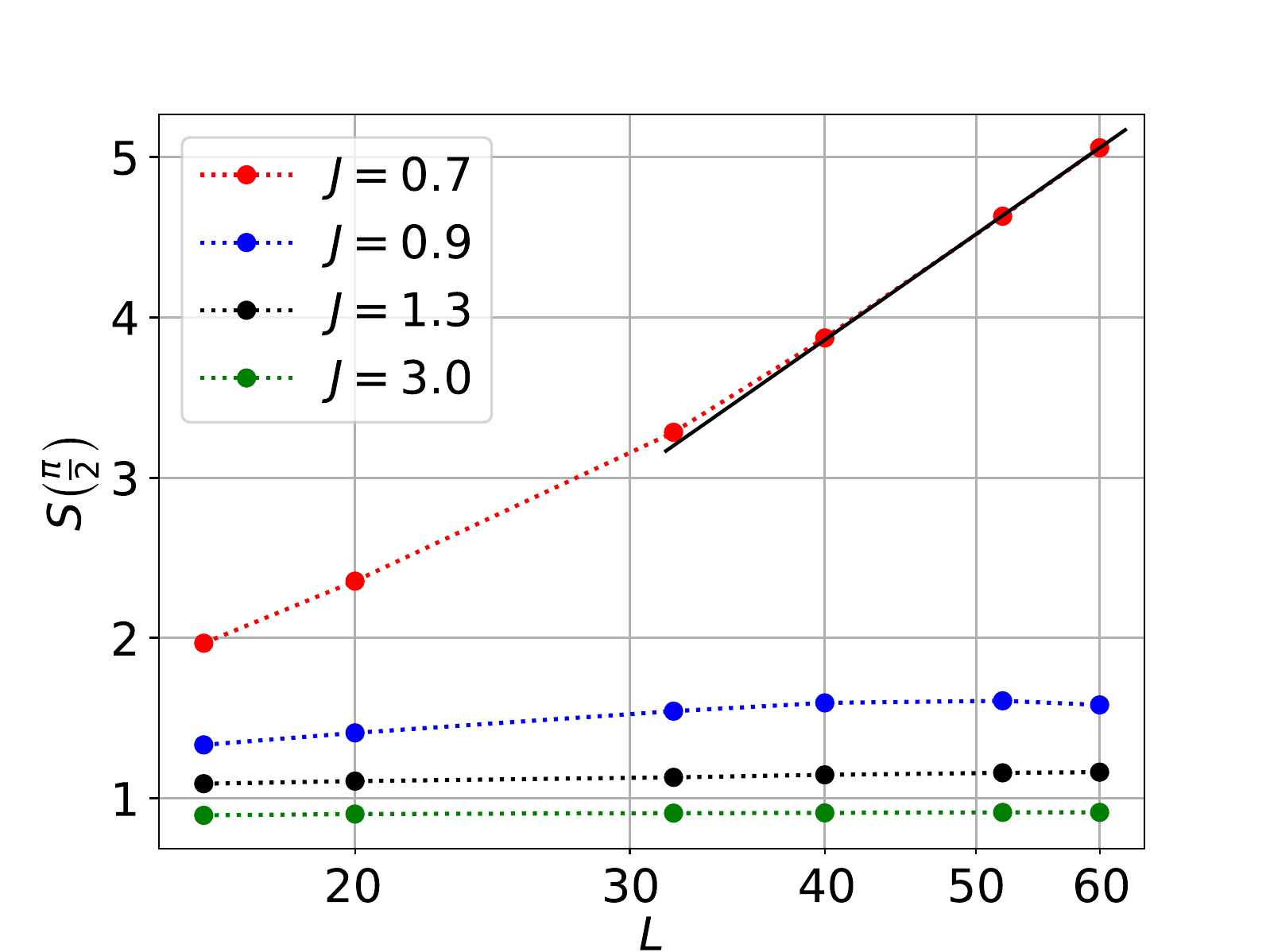}
\caption{
Spin-structure factor $S(Q)$ at $Q=\pi / 2$ for different $J$ as functions of $L$ on a logarithmic scale. 
}
\label{fig:div}
\end{figure}

This is supported by Fig.\ \ref{fig:div} where the $L$-dependence of the spin-structure factor at $Q=\pi / 2$ is shown for various interactions strengths $J$. 
One may expect \cite{MW66,Lia90,LC91,SS93,KM96} a logarithmic divergence for $L\to \infty$,
\be
  S(Q) \propto \ln^{1+\sigma} (L) 
  \: ,
\label{eq:log}  
\ee
corresponding to quasi-long-range order with an algebraic decay of the spin correlations at large distances $\Delta i$,
\be
  \langle \ff S_{i} \ff S_{i+\Delta i} \rangle 
  \propto 
  \frac
  {e^{- i Q \Delta i} \ln^{\sigma}(\Delta i)}
  {\Delta i}
  \: , 
\label{eq:inv}  
\ee
and with some exponent $\sigma$, describing logarithmic corrections to the $1/\Delta i$-behavior.

As is demonstrated with Fig.\ \ref{fig:div}, the $L$-dependence of $S(Q)$ is well described by Eq.\ (\ref{eq:log}) with $\sigma=0$ for $J < \jm$. 
Based on the present data, however, it does not seem possible to reach a reliable conclusion on the existence of logarithmic corrections and the value for the exponent $\sigma$. 
Fitting Eq.\ (\ref{eq:inv}) to our data for $\langle \ff S_{i} \ff S_{j} \rangle$ (see Fig.\ \ref{fig:distance}) turns out as even less predictive.
Calculations for substantially larger systems would be necessary.

For $J > \jm$ the spin-structure factor $S(Q)$ approaches a finite value in the limit $L\to \infty$, consistent with an exponential decay of the spin correlations.
The data are insufficient, however, to provide an accurate value of the critical interaction $\jm$ (see, however, Sec.\ \ref{sec:spingap} below).

\subsection{Discussion}
\label{sec:disc}

As the DMRG calculations (using the SU(2)-symmetric DMRG code) enforce a spin-isotropic ground state, ground-state expectation values of all observables must be invariant under SU(2) rotations.
This implies, for instance, $\langle \ff S_{i} \rangle=0$, also for $J<\jm$.
Hence, the sketch of the $90^{\circ}$ spin spiral with wavelength $\Delta i = 4$ that is provided by the arrows in Fig.\ \ref{fig:ladder} and in Fig.\ \ref{fig:chain}, is somewhat misleading and actually represents a classical picture only.

In addition, the quantum spin-spiral state is, in principle, also invariant under the full translation-symmetry group $\langle a \rangle$, generated by the primitive translation $a$.
This can be contrasted with the classical picture again (see Figs.\ \ref{fig:ladder} and \ref{fig:chain}), which would necessarily imply that the translational symmetry of the spin-spiral state was broken and that local observables, for example, must exhibit reduced symmetries described by the translation group $\langle 4a \rangle$.

Here, for the quantum case and for $J<\jd$, we have $\langle \ff S_{i+1} \ff S_{i+1+\Delta i} \rangle \ne \langle \ff S_{i} \ff S_{i+\Delta i} \rangle$ while $\langle \ff S_{i+2} \ff S_{i+2+\Delta i} \rangle = \langle \ff S_{i} \ff S_{i+\Delta i} \rangle$ for all $\Delta i$ (disregarding boundary effects), which reflects the symmetries described by the translation group $\langle 2a \rangle$.
This is interesting as this is not a generic feature of a quantum spin spiral but is instead explained by the manifest dimerization discussed in Sec.\ \ref{sec:short}:
Already the weak perturbation given by the presence of the chain boundaries is sufficient for $J<\jd$ to trigger the dimerization of the whole system and leads to reduced symmetries of the state, described by the translation-symmetry group $\langle 2a \rangle$.

The spin-dimerization transition would also manifest itself in the divergence $D(Q=\pi) \propto L \to \infty$ of the dimerization correlation function 
\begin{equation}
D(Q) \equiv \frac1L \sum_{ij}^{L} e^{iQ(R_{i}-R_{j})} \langle (\ff S_{i} \ff S_{i+1}) (\ff S_{j} \ff S_{j+1}) \rangle
\end{equation}
at $Q=\pi$. 
This is opposed to the weaker divergence $S(Q=\pi/2) \propto \ln L$ of the spin-correlation function Eq.\ (\ref{eq:scorr}). 
We note that the dimerization transition cannot be seen in $S(Q)$ and, vice versa, the magnetic transition cannot be seen in $D(Q)$. 

The spin dimerization is driven by the geometrical frustration as it, similar to the PKS mechanism, alleviates frustration and releases the according energy. 
As a side remark we note that it also affects the charge degrees of freedom, as is easily verified by analyzing the nearest-neighbor and the next-nearest-neighbor density-density correlations $n^{2} - \langle \ff n_{i} \ff n_{j} \rangle$ which show a similar, but much weaker pattern as the corresponding spin correlations displayed in the top panel of Fig.\ \ref{fig:shortj} with alternating stronger and weaker correlations on the $\diagup$ and $\diagdown$-bonds, respectively. 

Hence, avoiding frustration leads to a quantum spin-spiral order on the one hand and to spontaneous dimerization with breaking of translational symmetries on the other.
Since the representations of SU(2) spin rotations and of translations $\langle a \rangle$ in the Hilbert space commute, however, there is no reason to believe that generically the critical interaction be the same for both, dimerization and quasi-long-range spiral order. 
In fact, our data suggest $\jm < \jd$ (see Sec.\ \ref{sec:spingap} below for a precise determination of $\jm$).

\subsection{Classical correlations}

\begin{figure}[t]
\includegraphics[width=0.85\columnwidth]{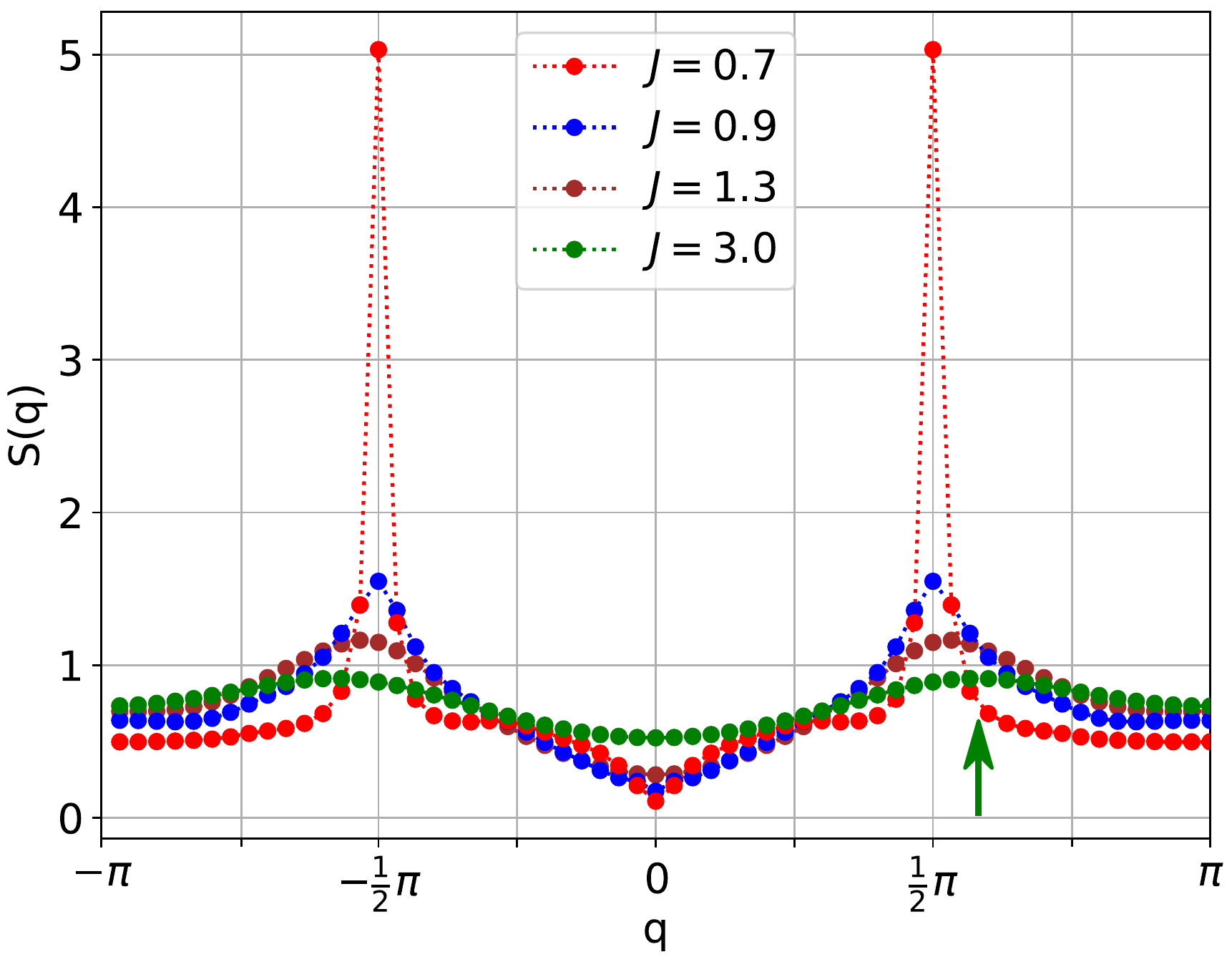}
\caption{
Spin-structure factor $S(q)$ for different $J$ as indicated and for $L=60$ sites. 
}
\label{fig:sofq}
\end{figure}

For $J \to \jm$ but $J>\jm$ the spin-structure factor $S(q)$ exhibits maxima at $Q_{\rm max.} = \pm \pi / 2$.
Fig.\ \ref{fig:overview} gives an overview, and Fig.\ \ref{fig:sofq} shows the detailed behavior.
Apparently, the maxima become more and more shallow with increasing $J$, as one would expect. 
But at the same time the position of the maxima crosses over to $Q_{\rm max.} \to \pm 1.82 \approx \pm 104.5^{\circ}$ for strong $J$, see the arrow in Fig.\ \ref{fig:sofq} for $J=3.0t$. 

This can easily be interpreted as a sign of {\em classical} spin correlations. 
To this end we assume that the magnetic degrees of freedom for strong $J$ are well described by an effective classical Heisenberg model with antiferromagnetic Heisenberg exchange $J_{\rm Heis.}$ which is nonzero between nearest neighbors on the zigzag ladder only. 
The classical ground state is a classical spin spiral with pitch angle $\vartheta$ between neighboring spins on a rung and $2\vartheta$ along the legs. \cite{Dub16}
For $\vartheta = 90^{\circ}$ we get the classical spin structure as sketched in Figs.\ \ref{fig:ladder} and \ref{fig:chain}.
Here, we determine the optimal pitch angle $\vartheta_{0}$ by minimizing the ground-state energy 
\be
  E_{0}(\vartheta) 
  = 
  J_{\rm Heis.} z_{\rm rung} L  S_{\rm cl.}^{2} \cos \vartheta
  +
  J_{\rm Heis.} z_{\rm leg} L S_{\rm cl.}^{2} \cos (2\vartheta) \: , 
\ee
where $S_{\rm cl.}$ is the length of the classical spin and $z_{\rm rung}=2$ and $z_{\rm leg}=2$ are the number of nearest neighbors along the rungs and the legs, respectively. 
This yields $\vartheta_{0} = \pm \arccos(-z_{\rm rung} / 4 z_{\rm leg}) = \pm 104.5^{\circ}$ and implies that $S(q)$ has maxima at $Q_{\rm max.}
= \vartheta_{0}$.
We conclude that for strong $J$ an effective classical Heisenberg model with couplings $J_{\rm Heis.}$ between neighboring spins correctly  
predicts the maximum position of $S(q)$.
Of course, the classical long-range spin-spiral ordering must be seen as an artifact in the strong-$J$ regime and reflects the missing quantum fluctuations.

\subsection{Spin gap}
\label{sec:spingap}

Let us finally discuss the spin gap $\Delta E_{\rm S}$, which is defined as the energy difference between the ground states in the invariant $S_{\rm tot}=1$ and $S_{\rm tot}=0$ subspaces: 
\be
  \Delta E_{\rm S} = E_0(S_{\rm tot}=1) - E_0(S_{\rm tot}=0) \: .
\ee
In the infinite-$J$ limit, the spin gap is given by $\Delta E_{\rm S} = J$ with perturbative corrections for finite but strong $J$ of the order of $t^{2}/J$; see Ref.\ \onlinecite{TSU97b} for a detailed discussion for the case of the one-dimensional (not frustrated) Kondo lattice.
For the unfrustrated case, the spin gap only closes in the $J\to 0$ limit, where a BCS-type behavior has been found in DMRG calculations:  \cite{SNUI96}
\be
  \ln (\Delta E_{\rm S}/t)
  \propto
  - \frac{t}{J} \: .
  \label{eq:sg} 
\ee
As compared to the Kondo-impurity case, it is enhanced, see Ref.\ \onlinecite{TSU97b} for a discussion.

\begin{figure}[t]
\includegraphics[width=0.8\columnwidth]{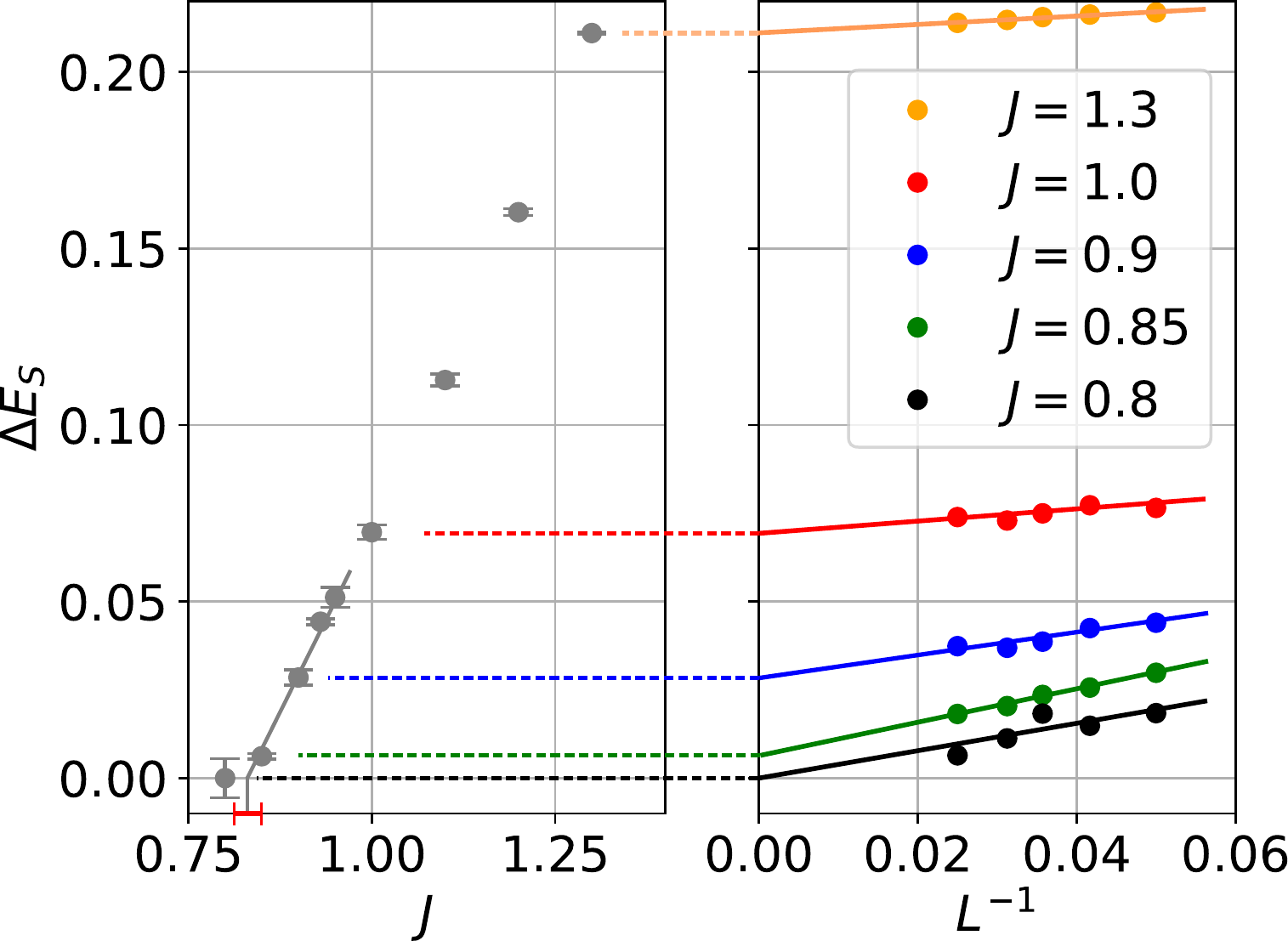}
\caption{
{\em Right:} 
Spin gap $\Delta E_{\rm S}(L)$ as a function of $1/L$ for different $J$ as indicated. 
Lines represent the results of a linear fits $\Delta E_{\rm S}(L) - \Delta E_{\rm S}(\infty) \propto 1/L$. 
{\em Left:} 
Values for $\Delta E_{\rm S}(\infty)$ obtained by extrapolation to the $1/L \to 0$ limit 
as a function of $J$. 
Error bars indicate the uncertainty of the extrapolation.
Fitting the $J$-dependence of $\Delta E_{\rm S}(\infty)$ (gray line) yields a critical value $\jm = \jmv \pm 0.03t$ (see horizontal red bar).
}
\label{fig:deltas}
\end{figure}

The right panel of Fig.\ \ref{fig:deltas} displays the $L$-dependence of the spin gap $\Delta E_{\rm S}(L)$ for various coupling strengths. 
For strong $J$ down to $J=1.3t$ (see figure), the spin gap is an almost perfectly linear function of $1/L$, and extrapolation to the infinite-system limit $1/L \to 0$ is straightforward. 
Extrapolated values for $\Delta E_{\rm S}(\infty)$ as a function of $J$ are shown in the left panel of the figure.
For smaller $J$ the data for $\Delta E_{\rm S}(L)$ somewhat scatter (Fig.\ \ref{fig:deltas}, right), and linear extrapolation to $1/L \to 0$ becomes slightly less predictive as indicated by the error bars on the extrapolated values for the spin gap (left).
However, a sizable error is found for $J=0.8t$ only.

The data are consistent with a continuous closure of the spin gap described by a linear $J$-dependence, 
\be
  \Delta E_{\rm S} \propto J-\jm
\: , 
\ee 
for $J\to \jm$. 
For the value of the critical coupling we find $\jm = \jmv \pm 0.03t$. 
The error is almost an order of magnitude larger than the error on the critical interaction $\jd$ for the dimerization. 
Note, however, that taking both errors into account, the data imply that $\jm < \jd$ (see also discussion in Sec.\ \ref{sec:disc}).

To crosscheck the presence of the magnetic phase transition and the value of the critical interaction, we have also analyzed the $J$-dependence of the spin-correlation length $\xi$ and of the spin-structure factor $S(Q)$. 
For a large but finite system size $L$, both observables appear to diverge with $J \to \jm$ at the same critical interaction $\jm \approx \jmv$.
A full finite-size scaling analysis based on data for various $L$ yields the same value for $\jm$.
The data and the corresponding discussion is given in Appendix \ref{sec:fss}.

Finally, one may easily extract the charge gap 
$\Delta_{\rm C} = [E_{0}(N+2) + E_{0}(N-2) - 2 E_{0}(N)] / 2$
from DMRG calculations for fillings $n=N/L$ slightly off half-filling ($n=1$).
As the spin gap, the charge gap $\Delta_{\rm C}$ is of the order of $J$ in the strong-$J$ regime.
It decreases with decreasing $J$ but stays finite at and below $\jm$. 
This is consistent with our expectation that the systems is an insulator for arbitrary $J>0$ at half-filling.

\section{Conclusions}
\label{sec:con}

Opposed to the conventional one-dimensional Kondo lattice, the Kondo lattice on the zigzag ladder, with equal hopping on the legs and rungs, exhibits at least two quantum-phase transitions at half-filling as a function of $J$.
The complex phase diagram arises from the competition between different mechanisms at work: 
(i) the Kondo screening or the formation of local Kondo correlations, 
(ii) the nonlocal, in most cases antiferromagnetic correlations due to exchange processes mediated by the conduction-electron system, and 
(iii) the frustration of magnetic order due to the geometry of the zigzag ladder.
Although DMRG calculations are tedious, particularly in the weak-$J$ regime, the zigzag Kondo ladder offers the unique possibility to study this competition at the numerically exact level. 

For strong $J$, the unique total-spin-singlet ground state is dominated by Kondo correlations, by very weak nonlocal correlations between the local spins, and by strongly gapped spin and charge excitations.
Even in this regime, however, there are clear signs for frustration of antiferromagnetic spin correlations since the spin-structure factor $S(q)$ (weakly) peaks at $q \approx 104.5^{\circ}$.
This is exactly the optimal pitch angle of the ground-state spin spiral in the corresponding classical zigzag Heisenberg model and thus indicates the presence of effective antiferromagnetic couplings between neighbors on the rungs and legs and therewith frustrated magnetism.
It is clear, though, that the classical compromise, i.e., spiral magnetic long-range order, is strongly suppressed by quantum fluctuations.

With decreasing $J$, the maximum of $S(q)$ grows and its position shifts towards $q=90^{\circ}$. 
Nonlocal antiferromagnetic correlations, at the cost of the local ones, become stronger, and the electronic degrees of freedom start to play a more important role.
Short-range correlations between the local spins but also between the local conduction-electron spins, triggered by the presence of the boundaries of the ladder, develop Friedel oscillations with wavelength $\Delta i = 4a$ which decay towards the center of the ladder.

For $J<\jd$ frustration is alleviated by spin dimerization, i.e., by spontaneous breaking of translational symmetry. 
This is indicated by a nonzero dimerization order parameter, defined as the asymmetry of spin correlations on geometrically equivalent rungs.
The transition at $\jd = \jdv \pm 0.005t$ is continuous or possibly weakly first order and preempts the transition to the magnetic state at lower $J$. 
This scenario is different but reminiscent of partial Kondo screening (PKS), an intermediate phase between a correlated insulator with dominant local Kondo correlations and gapped spin excitations on the one hand and a gapless magnetic state driven by indirect exchange interactions on the other.
PKS has been suggested for the periodic Kondo and Anderson models on frustrated two-dimensional lattices.
In the dimerized phase ferromagnetic correlations develop on every second rung such that the remaining nearest-neighbor bonds form a bipartite structure.
Dimerization, like PKS, thus assists the formation of antiferromagnetic order. 

Quasi-long-range order sets in at $J=\jm$, as indicated by the closure of the spin gap, the divergence of the spin-correlation length and the divergence of the spin-structure factor at $q=\pi/2$ for $J \to \jm$ ($J>\jm$). 
The critical interaction can be extracted consistently from all these observables. 
We find $\jm = \jmv \pm 0.03t$ based on data at finite but large $L$. 
This is corroborated by finite-size scaling analysis based on data for various system sizes.
Within the magnetic phase for $J<\jm$, we see a logarithmic divergence of $S(\pi/2)$ with increasing $L$.
While dimerization assists magnetic order, the magnetic transition is not immediate but takes place at a somewhat weaker coupling $\jm < \jd$.
We note that since the representations of SU(2) spin rotations and of translations in the Hilbert space commute, there is no reason to believe that generically the critical interaction be the same for both, dimerization and quasi-long-range spiral order.

The observed quasi-long-range order is a quantum spin spiral characterized by $Q_{\rm max}=\pi/2$. 
This can be made plausible by looking at the pattern of ferro- and antiferromagnetic nearest-neighbor correlations for $J<\jd$ (see, e.g., $J=0.87t$ in Fig.\ \ref{fig:shortj}).
The same pattern would be generated with an Ising-type long-range order: $...\uparrow, \uparrow, \downarrow, \downarrow, \uparrow, \uparrow, \downarrow, \downarrow ...$ that is {\em commensurate} with the dimerization and would be described by $Q_{\rm max}$.
However, quantum fluctuations must destroy any Ising-type state. 
The result is a quasi-long-range $90^{\circ}$ quantum spin spiral with dimerized short-range correlations, a state with no classical analog.

A well known paradigm of dimerization in frustrated quantum-spin models is given by the $J_{1}$-$J_{2}$ spin-$1/2$ Heisenberg chain at the  Majumdar-Ghosh point. \cite{MG69}
The ground state of the spin-dimerized phase is two-fold degenerate. 
Hence, the Lieb-Schultz-Mattis (LSM) theorem \cite{LSM61} requires that the spin excitations be gapped. 
This is opposed to the one-dimensional Kondo lattice studied here: 
Disregarding charge fluctuations, one could argue that at half-filling there is a local spin-$1/2$ per site in the conduction-electron system. 
This implies that for both, a transitionally invariant or a dimerized ground state, the respective unit cell would have an integer spin.
Following Haldane's conjecture, \cite{Hal83} one would therefore rather expect a gapped phase on both sides of the spin-dimerization transition, and this is found here in fact.
However, the transition to a gapless magnetic state at $\jm$ (within the dimerized phase) demonstrates that charge fluctuations and the itineracy of the conduction electrons play an essential role.
We conclude that the physics found here is not covered by the Majumdar-Ghosh paradigm.

A spin-dimerized phase in the one-dimensional Kondo lattice with nearest-neighbor hopping only has been observed previously at {\em quarter-filling}. \cite{XPMA03,XM08}
In these studies, dimerization is found at weak $J$ and is explained as being induced by effective nearest-neighbor and next-nearest-neighbor RKKY couplings, i.e., an effective low-energy Heisenberg model applies, and charge fluctuations in the electronic system become irrelevant for the low-energy physics.
For the calculated effective Heisenberg couplings, a dimerized Majumdar-Gosh-like ground state emerges.
The LSM theorem applies since, at quarter-filling, the unit cell of the dimerized state contains a half-integer total spin, and a gapless spin-excitation spectrum has been found. 
On the contrary, the phase transitions found here, for the {\em half-filled} Kondo lattice, are of different origin, since even for the low-energy excitations, charge fluctuations cannot be disregarded.
Namely, RKKY perturbation theory simply does not apply to the numerically studied regime of exchange couplings. 
The computed nearest-neighbor ($J^{\rm RKKY}_{1}$) and next-nearest-neighbor ($J^{\rm RKKY}_{2}$) RKKY couplings are at variance with the DMRG data for the weakest couplings accessible to the numerics.
Only for considerably weaker $J$, do we expect a spin-only Heisenberg picture to apply, albeit featuring different physics since 
$J^{\rm RKKY}_{1}>0$ (antiferromagnetic) and next-nearest-neighbor $J^{\rm RKKY}_{2}<0$ (ferromagnetic) are not frustrated. 

Concluding, the observed complex competition of various mechanisms and the resulting quantum-phase transitions take place in the intermediate-$J$ regime and thus exclude a simple explanation in terms of simple effective low-energy models.
Future studies could address the model in different parameter regimes, such as next-nearest-neighbor hopping $t' \ne t$ and fillings $n$ away from half-filling. 
Particularly for the challenging regime of fillings off but close to half-filling, it would be interesting to see if the phase diagram gets even more involved.
Magnetic frustration in one-dimensional systems has also been studied in related models, such as the simpler $J_{1}$-$J_{2}$ Heisenberg chain (see Ref.\ \onlinecite{KSSR10} and references therein) or, using the Bethe ansatz, in the $t$-$J$ model at the supersymmetric point with a finite concentration of impurities which give rise to frustration, but do not break integrability, \cite{SZ97} or in supersymmetric $t$-$J$ chains with next-nearest-neighbor interactions. \cite{ZKZ01} 
A direct comparison between nonperturbative numerical and analytical methods can be highly instructive, particularly with respect to the low-lying excitations. \cite{RP17}
Those questions, however, are beyond the present study, but will be pursued in forthcoming work.
\\

\acknowledgments
This work has been supported by the Deutsche Forschungsgemeinschaft through the Sonderforschungsbereich 668 (project A14).
Calculations have been performed at the PHYSnet Computing Center at the University of Hamburg.

\appendix

\section{Finite-size scaling}
\label{sec:fss}

To further support the existence of the magnetic phase transition and to crosscheck the value for the critical inteaction $\jm$, we discuss some more data for the $J$-dependence of the correlation length and of the spin-structure factor at $q=\pi/2$ at various system sizes up to $L=40$.

\begin{figure}[t]
\includegraphics[width=0.8\columnwidth]{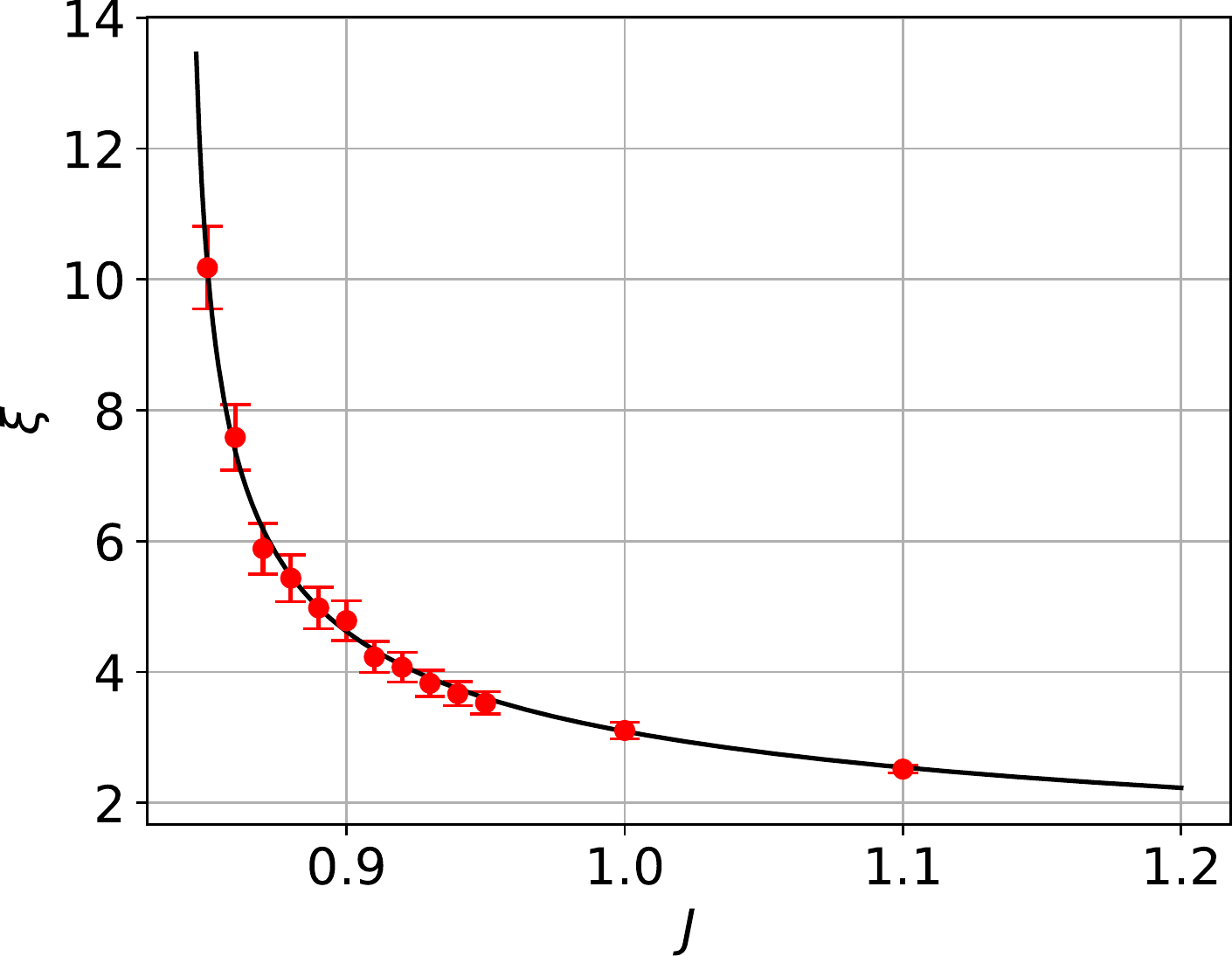}
\caption{
Correlation length $\xi$ as function of $J$ (for $J>\jm$) at $L=40$.
Error bars on $\xi$ result from fitting the DMRG data for the distance dependence of the spin correlation function $\langle \ff S_{i} \ff S_{j} \rangle$.
Solid line: data for $\xi$ are fitted by $\xi = \mbox{const} \times (J-\jm)^{-\nu}$. 
Optimal values: $\jm \approx 0.84t$, $\nu \approx 0.4$, $\mbox{const} \approx 1.48$.
}
\label{fig:corr}
\end{figure}

\begin{figure}[b]
\includegraphics[width=0.8\columnwidth]{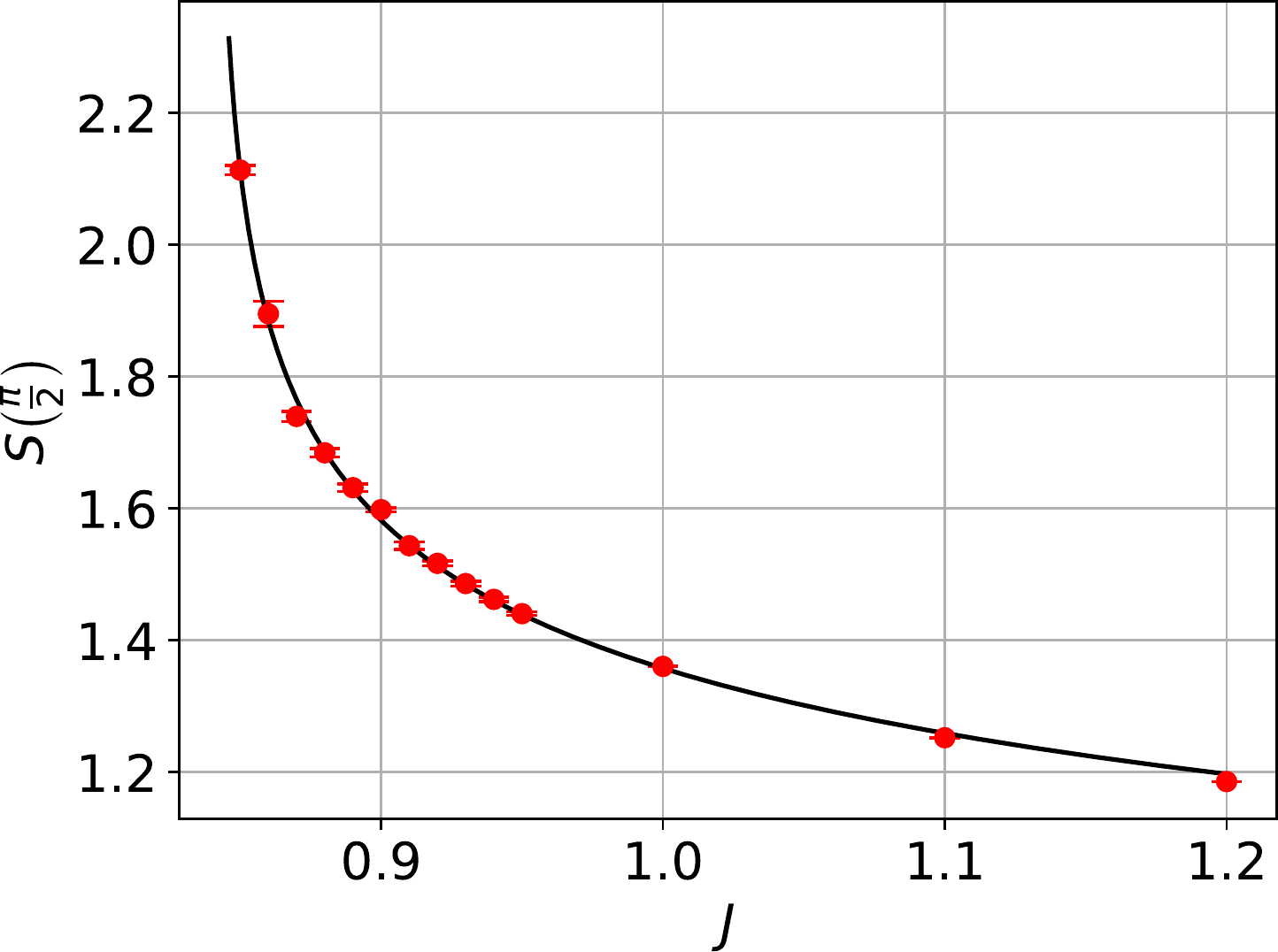}
\caption{
$J$-dependence (for $J>\jm$) of the spin-structure factor $S(q)$ at wave vector $q=\pi/2$ and $L=40$.
Error bars reflect the scattering of the data for different bond dimensions.
Data are fitted by $S(\pi/2)=\mbox{const} \times (J-\jm)^{-\gamma}$. 
Optimal values: $\jm \approx 0.84t$, $\gamma \approx 0.15$, $\mbox{const} \approx 1.02$.
}
\label{fig:sus}
\end{figure}

The correlation length $\xi$ is obtained by fitting the data for the distance dependence of the spin correlation function to $\langle \ff S_{i} \ff S_{j} \rangle = \mbox{const} \times e^{- |i-j| / \xi}$ in the nonmagnetic gapped phase for $J > \jm$.
Fig.\ \ref{fig:corr} displays $\xi$ as a function of the coupling $J$ at $L=40$. 
We find the expected power-law behavior close to the magnetic transition, and $\xi$ appears to diverge at $\jm \approx \jmv$.
This value is consistent with the results for the spin gap (see the discussion of Fig.\ \ref{fig:deltas}).
One should note that $\xi$ is obtained for the ``homogeneous'' ground state if $J > \jd \approx \jdv$, while the calculations for $J<\jd$ are done for the dimerized ground state.
This explains the slightly irregular $J$-dependence of $\xi$ close to $\jd \approx \jdv$.

Fig.\ \ref{fig:sus} displays the DMRG data for the spin-structure factor at $q=\pi/2$ and for $L=40$. 
For its $J$-dependence close to $\jm$, we find $S(\pi/2)$ to follow a power law, as expected. 
It appears to diverge at the same critical interaction strength $\jm \approx \jmv$.
Again, we also note a slight deviation from the power-law behavior close to the spin-dimerization transition $\jd$. 

Clearly, truly divergent behavior can only occur in the thermodynamic limit $L\to \infty$.
Hence, for a refined determination of the location of the magnetic transition $\jm$, a finite-size scaling analysis is helpful: \cite{fss}
Assuming that the transition to the magnetic state is continuous, one would expect $\xi \propto (J-\jm)^{-\nu}$ close to $\jm$ ($J> \jm$).
In addition, we have $\chi_{q} \propto (J-\jm)^{-\gamma}$ for the critical behavior of the static magnetic susceptibility 
$\chi_{q} = \lim_{\beta \to \infty} L^{-1}\sum_{ij} e^{-iq|i-j|} \int_{0}^{\beta} d\tau \langle S_{i}(\tau) S_{j}(0)) \rangle$.
The standard finite-size scaling ansatz for the $L$-dependence of the susceptibility is given by 
$\chi_{q}(L) = \xi^{\gamma/\nu} f(L / \xi)$ where $f$ is the scaling function of the dimensionless ratio $x=L/\xi$.
This implies 
\begin{equation}
\chi_{q}(L) = L^{\gamma/\nu} \tilde{f}(L^{1/\nu} (J-\jm))
\end{equation}
with $\tilde{f}(x)=x^{-\gamma}f(x^{\nu})$.

\begin{figure}[t]
\includegraphics[width=0.8\columnwidth]{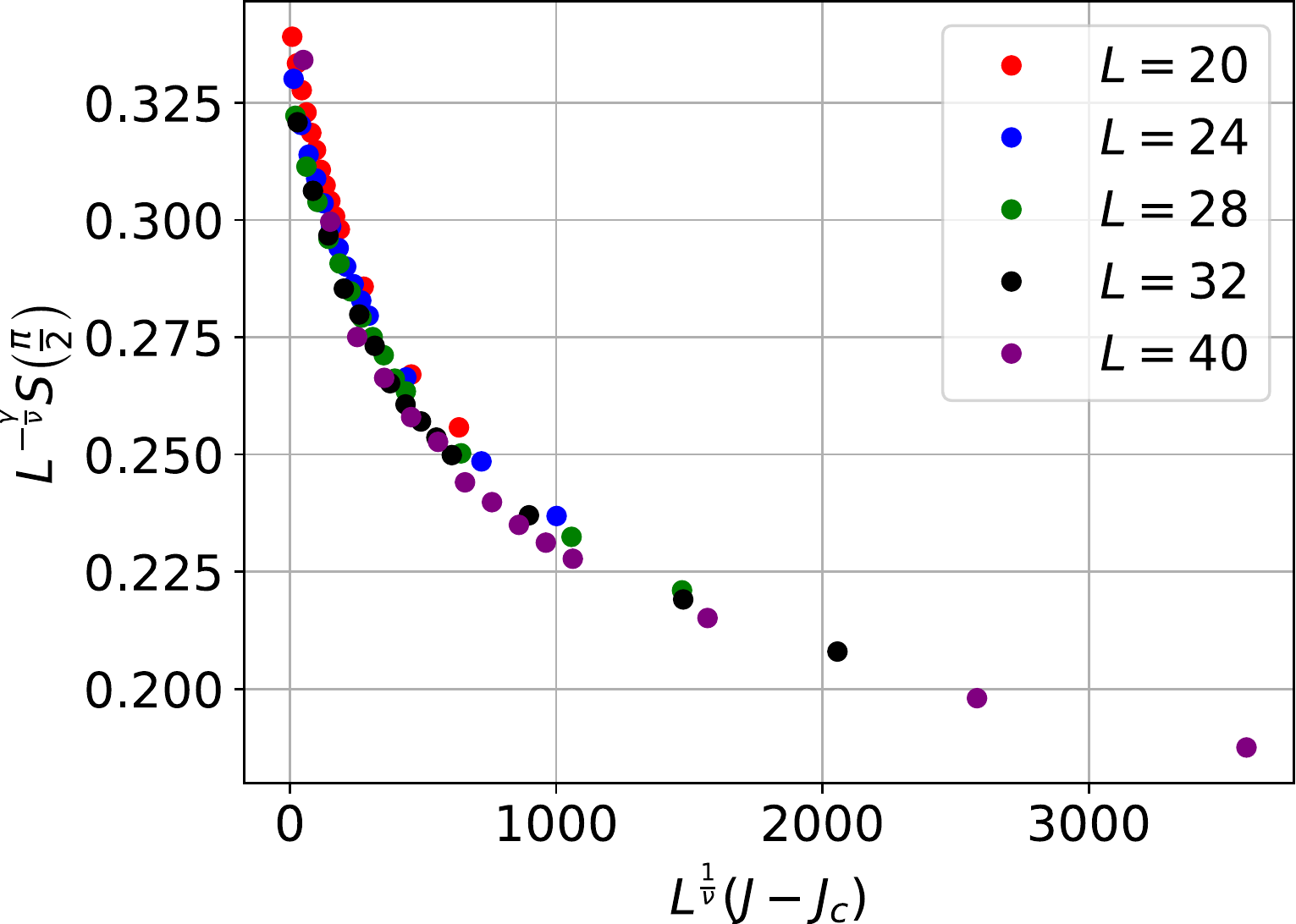}
\caption{
Finite-size scaling for $S(\pi/2)$, see text.
The data for different $L$ as indicated and for $J=0.85, 0.86, ..., 0.95$ collapse to a single line at $\jm \approx 0.845 t$, $\nu \approx 0.4$, $\gamma \approx 0.2$.
}
\label{fig:coll}
\end{figure}

In Fig.\ \ref{fig:coll} we plot $L^{-\gamma/\nu} \, S(\pi/2)$ against $L^{1/\nu} (J - \jm)$ for various system sizes $L=20, 24, 28, 32, 40$.
All data collapse to the universal scaling function for $\jm \approx 0.844t$ and for the exponents $\nu \approx 0.4$ and $\gamma \approx 0.2$.
While the critical interaction $\jm$ can be determined rather precisely, up to a few per cent, there is a comparatively large error in the values of $\nu$ and $\gamma$, resulting mainly from the nearby spin-dimerization transition. 

\end{document}